\documentclass[12pt,preprint]{aastex}

\shorttitle{TESTING THE ACCRETION FLOW FOR SGR A*}
\shortauthors{Huang et al.}

\begin{document}


\title{TESTING THE ACCRETION FLOW WITH PLASMA WAVE HEATING MECHANISM FOR SAGITTARIUS A* BY THE 1.3MM VLBI MEASUREMENTS}


\author{Lei Huang\altaffilmark{1,2,3}, Rohta Takahashi\altaffilmark{4}, Zhi-Qiang Shen\altaffilmark{2}}

\altaffiltext{1}{Key Laboratory for Research in Galaxies and
Cosmology, The University of Sciences and Technology of China,
Chinese Academy of Sciences, Hefei 230026, China;
mlhuang@ustc.edu.cn}

\altaffiltext{2}{Key Laboratory for Research in Galaxies and
Cosmology, Shanghai Astronomical Observatory, Chinese Academy of
Sciences, Shanghai 200030, China}

\altaffiltext{3}{Academia Sinica, Institute of Astronomy and
Astrophysics, Taipei 106, Taiwan}

\altaffiltext{4}{The Institute of Physical and Chemical Research,
2-1 Hirosawa, Wako, Saitama 351-0198, Japan}


\begin{abstract}

The vicinity of the supermassive black hole associated with the
compact radio source Sagittarius (Sgr) A* is believed to dominate
the observed emission at wavelengths near and shorter than $\sim$ 1
millimeter. We show that a general relativistic accretion flow,
heated via the plasma wave heating mechanism, is consistent with the
polarization and recent mm-VLBI observations of Sgr A* for an
inclination angle of $\sim 45^\circ$, position angle of $\sim
140^\circ$, and spin $\lesssim 0.9$. Structure in visibilities
produced by the black hole shadow can potentially be observed by 1.3
mm-VLBI on the existing Hawaii-CARMA and Hawaii-SMT baselines. We
also consider eight additional potential mm-VLBI stations, including
sites in Chile and New Zealand, finding that with these the basic
geometry of the emission region can be reliably estimated.

\end{abstract}


\section{Introduction}
\label{intro}

The compact radio source Sagittarius (Sgr) A*, residing at the
Galactic center, is believed to be the best candidate of
suppermassive black hole \citep{Scho02,Ghez05}. Based on
observations of the orbiting stars, the mass of its central object
is measured as over 4 million $M_\odot$ \citep{Ghez08,Gill09}.
However, it remains to be conclusively proved that Sgr A* is a black
hole, and the explanations of the broad-band emission are even
controversial. Various morphology of the emission region, torus,
jet, and their combinations, are provided by different models. The
observations at millimeter or longer wavelengths cannot distinguish
them due to the interstellar scattering. However, the interstellar
scattering is becoming less severe as the observational wavelength
decreases to millimeter band, so that the intrinsic size of Sgr A*
has been measured at 7mm and 3mm \citep{Bowe04, Shen05}. According
to the extrapolated scattering size, the intrinsic structure of Sgr
A* would be dominant, i.e., the sub-millimeter VLBI promises to
produce unadulterated images of its intrinsic structure. However,
lack of baselines results in poor uv-coverage, making the imaging
impossible. Thus, theoretical modelling of Sgr A* will be critical
to both planning and interpreting sub-millimeter VLBI observaions
\citep{Huang07}.

Recently, \citet{Doel08} reported observations at 1.3 mm with a
very-long-baseline interferometry (VLBI) array consisting of the
Arizona Radio Observatory Sub-millimeter Telescope (SMT), Combined
Array for Research in Millimeter-wave Astronomy (CARMA), and James
Clerk Maxwell Telescope (JCMT). Robust detections are obtained on
baseline CARMA-SMT at $\sim$ 500 km and baseline JCMT-SMT at
$\sim$ 4500 km. An upper limit is yielded on baseline CARMA-JCMT
at $\sim$ 3000 km. Since these visibility measurements are too few
to produce an image, a circular Gaussian modelling results in a
full width at half maximum (FWHM) of $\sim 37^{+16}_{-10} \mu {\rm
as}$, with the scattering effects removed \citep{Doel08}. However,
also pointed out by them, this size is somewhat unexpected,
because it is smaller than the 'black hole shadow' size of $\sim
50 \mu {\rm as}$ regardless of the black hole spin \citep{FMA00}.
It might be possible that the emission really comes from the
inside of black hole shadow. On the other hand, a structure with a
central black hole shadow itself, deviating from the Gaussian,
might be implied by the visibility measurements.

In this paper, we present a general relativistic accretion flow
model with plasma wave heating mechanism (Sec.\ref{mod}). The
polarization observations can be well explained with appropriate
viewing angle and other model parameters set for arbitrary black
hole spin. We then tend to use eight potential stations for future
1.3mm VLBI measurements to perform visibility analysis
(Sec.\ref{cov}). We perform a visibility analysis using eight
potential stations for future 1.3 mm VLBI measurements. We test the
images with recent 1.3 mm VLBI measurements and investigate the
implications for the properties of the black hole and accetion flow
(Sec.\ref{vis}). Comparisons to the earlier work and evaluations on
the potential $(u,v)$ coverage are given in the final Section.

\section{General Relativistic Keplerian Accretion Flow}
\label{mod}

We follow the magneto-rotational-instability (MRI) driven accretion
flow \citep{MLC01,Liu07} where the primary mechanism for generation
of turbulence and viscous stress in accretion flows supported by
magnetohydrodynamics (MHD) simulations \citep{BH91,BH98}. When
deriving the basic equations of the general relativistic accretion
flow, we use the energy momentum tensor $T^{\mu\nu}$ given as
$T^{\mu\nu} = n(m_p+m_e) \eta u^\mu u^\nu + p g^{\mu\nu} +
t^{\mu\nu} + q^\mu u^\nu + q^\nu u^\mu$ where $n$, $u^\mu$, $q^\mu$,
$t^{\mu\nu}$, $p$, $\eta$, $m_p$ and $m_e$ are the rest number
density, four-velocity, heat-flux four-vector, viscous tensor,
pressure, relativistic enthalpy, the mass of the proton and the mass
of the electron, respectively. The basic equations for the
relativistic hydrodynamics are then derived from the baryon-mass
conservation $\nabla_\mu (n u^\mu)=0$, the energy-momentum
conservation $\nabla_\nu T^{\mu\nu}=0$ and an equation of state.
Detailed derivations can be found in \citet{Taka07}. In this study,
the Boyer-Lindquist coordinate is used for the description of the
Kerr metric.

Following \citet{Liu07} and \citet{Huang08}, we introduce model parameters as
the ratio of the total stress to the magnetic field energy density $\beta_\nu$,
the ratio of the magnetic field energy density to the gas pressure $\beta_p$,
dimensionless constant for electron heating rate $C_1$, constant mass accretion rate
$\dot{M}$, inclination angle $i$, and position angle $\Theta$. An additional
parameter of black hole spin $a$ is also introduced.
The mean black hole mass of $4.1 \times 10^6 M_\odot$, measured from the orbit of the
short-period star SO-2 for the distance of $8.0$ kpc \citep{Ghez08}, is used.
We adopt the simplified form for the viscous tensor component $t^r_{~\phi}$ \citep{Laso94,Abra96}
\begin{eqnarray}
    t^r_\phi &=& - n(m_p+m_e) \nu \frac{A^{3/2} \Delta^{1/2} \gamma^3}{r^5}  d\Omega/dr,
\end{eqnarray}
where $A = r^4 + r^2 a^2 + 2Mra^2$, $\Delta=r^2 - 2Mr + a^2$, $\gamma$ is the
Lorentz factor, $\Omega = u^\phi /u^t = d\phi /dt$ is the angular velocity of the
accretion flow, and $\nu$ is the kinematic viscosity coefficient.
We then assume the $\alpha$-viscosity as
\begin{eqnarray}
    \nu &=& \frac{\beta_\nu\beta_p p}{n(m_p+m_e) (-r {\rm d}\Omega /{\rm d}r)},
\end{eqnarray}
where the product of $\beta_\nu$ and $\beta_p$ represents the $\alpha$ coefficient.

Instead of assuming a Keplerian accretion flow co-rotating with the
central black hole as was done in previous work, we include the
radial momentum conservation in the equation set. I.e., at any
radius, the radial velocity is solved by $\nabla_\mu T^{r\mu}=0$.
The angular velocity is a little lower than the Keplerian velocity
outside about ten gravitational radii to the center. In small radii,
the radial velocity increases significantly and the angular velocity
becomes much lower than the Keplerian velocity. The accreted plasma
passes through a sonic point at a radius inside the innermost stable
circular orbit, becomes super-sonic and finally crosses the event
horizon of the black hole at a speed of light. We mention that we
don't consider the causality of the accretion flow
\citep{GP98,Taka07}. We also adopt the viscous tensor component
derived from thin disk and the $\alpha$-viscosity coefficient for
simplicity. Such treatments are enough to generate rough dynamical
structure for the accretion flow, which can be used for
observational predictions at 1.3mm wavelength. For future
observations at much shorter wavelengths and with higher resolution,
causal viscosity should be calculated carefully to make more
realistic images of the vicinity of the black hole.

Here, the viscous heating rate is assumed to be equal to the
turbulence cascade rate, $\Gamma_{\rm vis}^+ \sim c_s B^2 /8\pi r
H_\theta$, where $B$ is the magnetic field amplitude determined by
the gas pressure and assumed $\beta_p$, $c_s$ is the sound speed,
and $H_\theta$ is the angular half-thickness of the disk. The
radiative cooling is considered for electrons only and the  viscous
heating rate for electrons is assumed to be different from that for
protons, and the radiative cooling for electrons is much more
effective. We adopt the heating mechanism by turbulent plasma wave
\citep{BE87,Liu06} for electrons as $\Gamma_{\rm acc}^+ = \alpha_e
k_B T_e \tau_{\rm acc}^{-1}$, where the timescale $\tau_{\rm acc} =
3 C_1 rH_\theta \bar{v}_e /c_s^2$, $\alpha_e = x[(3K_3(x) +
K_1(x))/4K_2(x) - 1]$ with $K_i(x)$ as the $i$th order modified
Bessel function of dimensionless temperature $x$, and $\bar{v}_e$ is
the mean electron speed. Therefore the energy equation for electrons
becomes
\begin{eqnarray}
    n u^r \alpha_e k_B \frac{{\rm d} T_e}{{\rm d} r} &=& \Gamma_{\rm acc}^+ + \Gamma_{ie}^+ - \Lambda_{\rm rad}^-, \label{energyeqne}
\end{eqnarray}
and the energy equation for both protons and electrons derived from local energy conservation
$\nabla_\mu T^{t\mu}=0$ is
\begin{equation}
n u^r \frac{d}{dr}[\eta {\mathcal E} (m_p + m_e)] = \Gamma_{\rm vis}^+ - \Lambda_{\rm rad}^-,
\label{energyeqn}
\end{equation}
where $\Gamma_{ie}^+$ is the Coulomb energy exchange rate between
electrons and protons and $\Lambda_{\rm rad}^-$ is the radiative
cooling rate calculated from the sum of the energy losses due to the
synchrotron radiation, the synchrotron self-Comptonization of the
soft photons and the bremsstrahlung. The temperatures of protons and
electrons can be solved with the Eq.\ref{energyeqn} and
Eq.\ref{energyeqne} combined.

The dynamical structure of the accretion flow is determined by
solving all the physical quantities on the equatorial plane and the
scale height from the outermost radius of $10^4$ Schwarzschild radii
to the event horizon. We assume homogeneous distribution in the
vertical direction at any radius and a configuration of magnetic
field to be parallel to the velocity field in the upper side of the
accretion flow and reversed in the lower side, which forms from the
initial field lines in vertical structure and the shearing of the
Keplerian accretion flow. Considering synchrotron emission of only
thermal population of electrons together with self-absorption and
birefringence effects, full general relativistic radiative transfer
along ray trajectories is then performed to give polarized images at
any given frequency. Corresponding equations can be found in
\citet{Huang08} and \citet{Huang09}.

The model parameters are set in the following procedure. First of
all, $\beta_\nu$ is fixed to be 0.7 when the turbulence saturates
\citep{PCP06}. Secondly, the inclination angle $i$ is fixed to be
$45^\circ$ as a typical value to reproduce high linear polarization
in sub-millimeter band \citep{Liu07,Huang08}. The position angle
$\Theta$ is chosen as $140^\circ$, which can make the accretion flow
reproduce the observed position angle of electric vector (EVPA) from
millimeter to near-infrared band \citep{Bowe05,Macq06,Marr07,Meye07}
with the mean external rotation measure of $\sim 5.6 \times 10^5
{\rm rad} \cdot {\rm m}^{-2}$ adopted \citep{Marr07}. Next, we
arbitrarily set a value for spin $a$ and choose an appropriate value
for $\beta_p$. In practice, if $a$ is set, $\beta_p$ should be small
enough to keep very low circular polarizations in the sub-millimeter
band reported by \citet{Marr06}. Finally, we adjust values for
$\dot{M}$ and $C_1$ and calculate the synchrotron radiation of
thermal electrons to reproduce the linear polarization degree of
$\sim 10\%$ in sub-millimeter band \citep{Aitk00} and overall
spectrum from centimeter to near-infrared band, fixing the flux
density at 1.3 mm to 2.4 Jy. In practice, the linear polarization is
most sensitive to $\dot{M}$ and the flux density is affected by the
combination of $\dot{M}$, $C_1$, and $\beta_p$.

\begin{figure}[h]
\vspace{-0mm}
\begin{center}
\includegraphics[width=14.0cm]{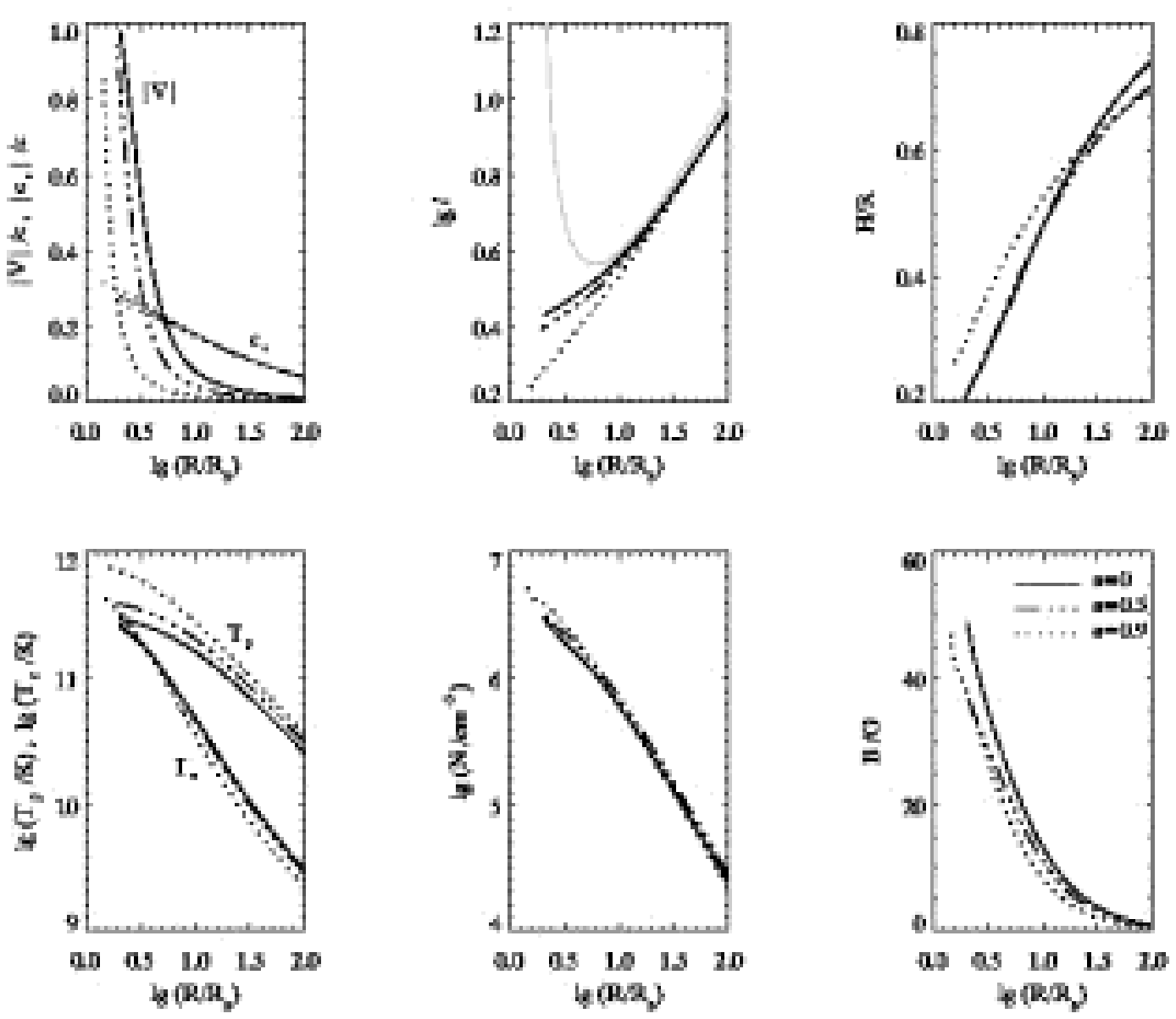}
\vspace{-5mm}\caption{
Global solutions of the accretion flow with plasma wave heating mechanism.
The solid lines represent the results for $a=0, \beta_p=0.4, C_1=0.202, \dot{M}=5 \times 10^{17} {\rm g \cdot s^{-1}}$. The dash-three-dotted lines are for $a=0.5, \beta_p=0.2, C_1=0.357, \dot{M}=3 \times 10^{17} {\rm g \cdot s^{-1}}$. The dotted lines are for $a=0.9, \beta_p=0.1, C_1=0.765, \dot{M}=1.2 \times 10^{17} {\rm g \cdot s^{-1}}$.
\textit{Top-left}: radial velocity and sound speed; \textit{Top-middle}: angular momentum, with the Keplerian one in solid grey line; \textit{Top-right}: ratio of scale height to radius;
\textit{Bottom-left}: temperatures of protons and electrons; \textit{Bottom-middle}: number density of electrons; \textit{Bottom-right}: amplitude of magnetic field.
\label{dynam}}
\end{center}
\end{figure}

\begin{figure}[h]
\vspace{-0mm}
\begin{center}
\includegraphics[width=14.0cm]{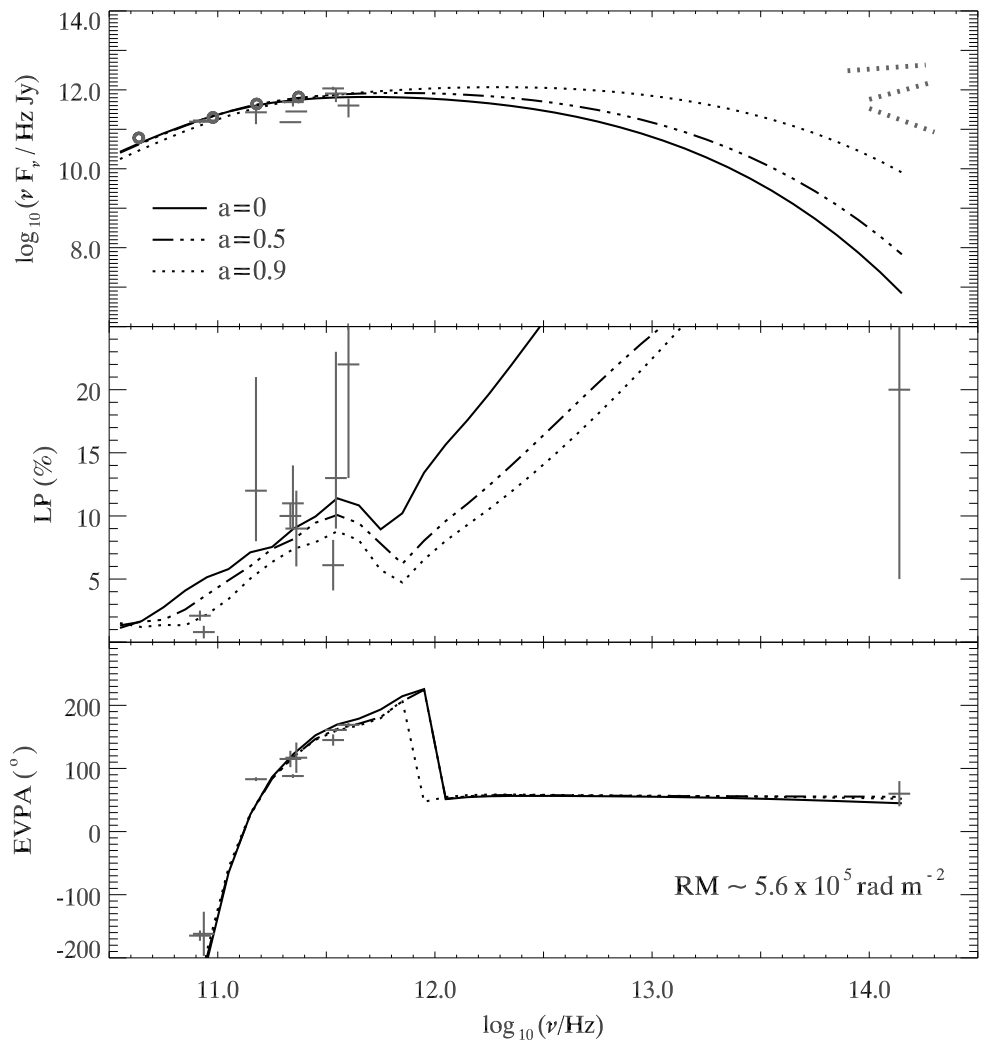}
\vspace{-5mm}\caption{
Spectrum of three accretion flow models with dynamical structures shown in Fig.\ref{dynam},  $i=45^\circ$, and $\Theta=140^\circ$. \textit{Top}: spectrum of total flux density of synchrotron emission; \textit{Middle}: linear polarization degree of synchrotron emission; \textit{Bottom}: position angle of electric vector (EVPA) with an external rotation measure of
$\sim 5.6 \times 10^5 {\rm rad} \cdot {\rm m}^{-2}$ considered.
\label{spec}}
\end{center}
\end{figure}

We adopt $i=45^\circ, \Theta=140^\circ$, which can explain the
polarization observations well, as our fiducial model, then choose
three sets of model parameters corresponding to three different
spins. In detail, those parameters are $a=0, \beta_p=0.4,
C_1=0.202, \dot{M}=5 \times 10^{17} {\rm g \cdot s^{-1}}$ for a
non-rotating black hole, $a=0.5, \beta_p=0.2, C_1=0.357, \dot{M}=3
\times 10^{17} {\rm g \cdot s^{-1}}$ for a mildly-rotating black
hole, and $a=0.9, \beta_p=0.1, C_1=0.765, \dot{M}=1.2 \times
10^{17} {\rm g \cdot s^{-1}}$ for a fast-rotating black hole. We
show the global solutions of the three cases in Fig.\ref{dynam}
and the spectra predicted by them in Fig.\ref{spec}. The observed
emission bump and high linear polarization degree in the
sub-millimeter band are predicted by the accretion flow. Another
jet/outflow component \citep{Liu07}, which is not shown in these
figures, contributes to the emission and depolarization in
millimeter and longer wavelengths. For the wavelength of 1.3mm we
are interested in, we assume the emission is dominated by the
accretion flow and the jet/outflow component may produce a weak
depolarization and a certain change in its EVPA.

\section{Potential $(u,v)$ Coverage For 1.3MM VLBI}
\label{cov}

\begin{figure}[h]
\vspace{-0mm}
\begin{center}
\includegraphics[width=7.5cm]{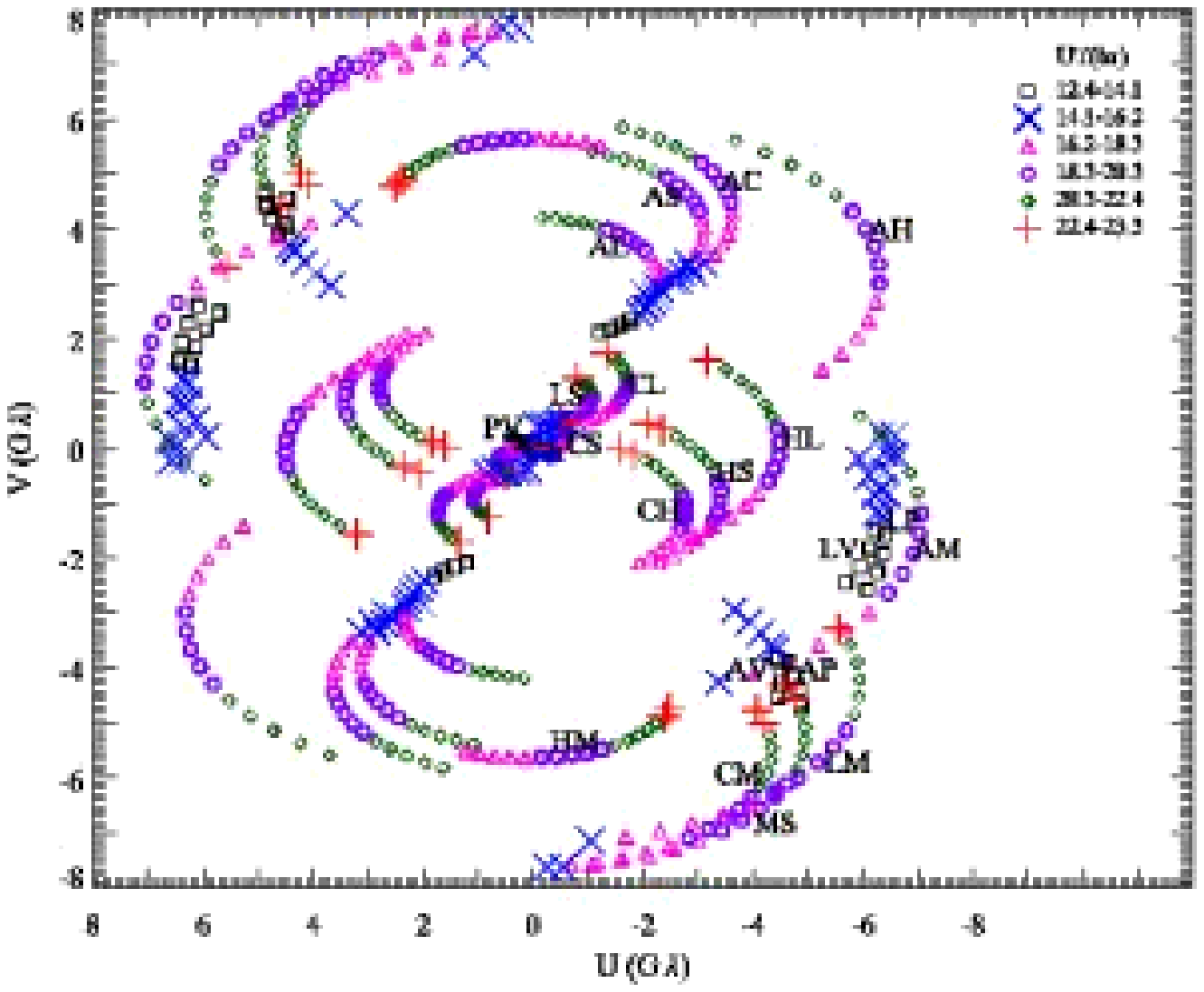}
\includegraphics[width=7.5cm]{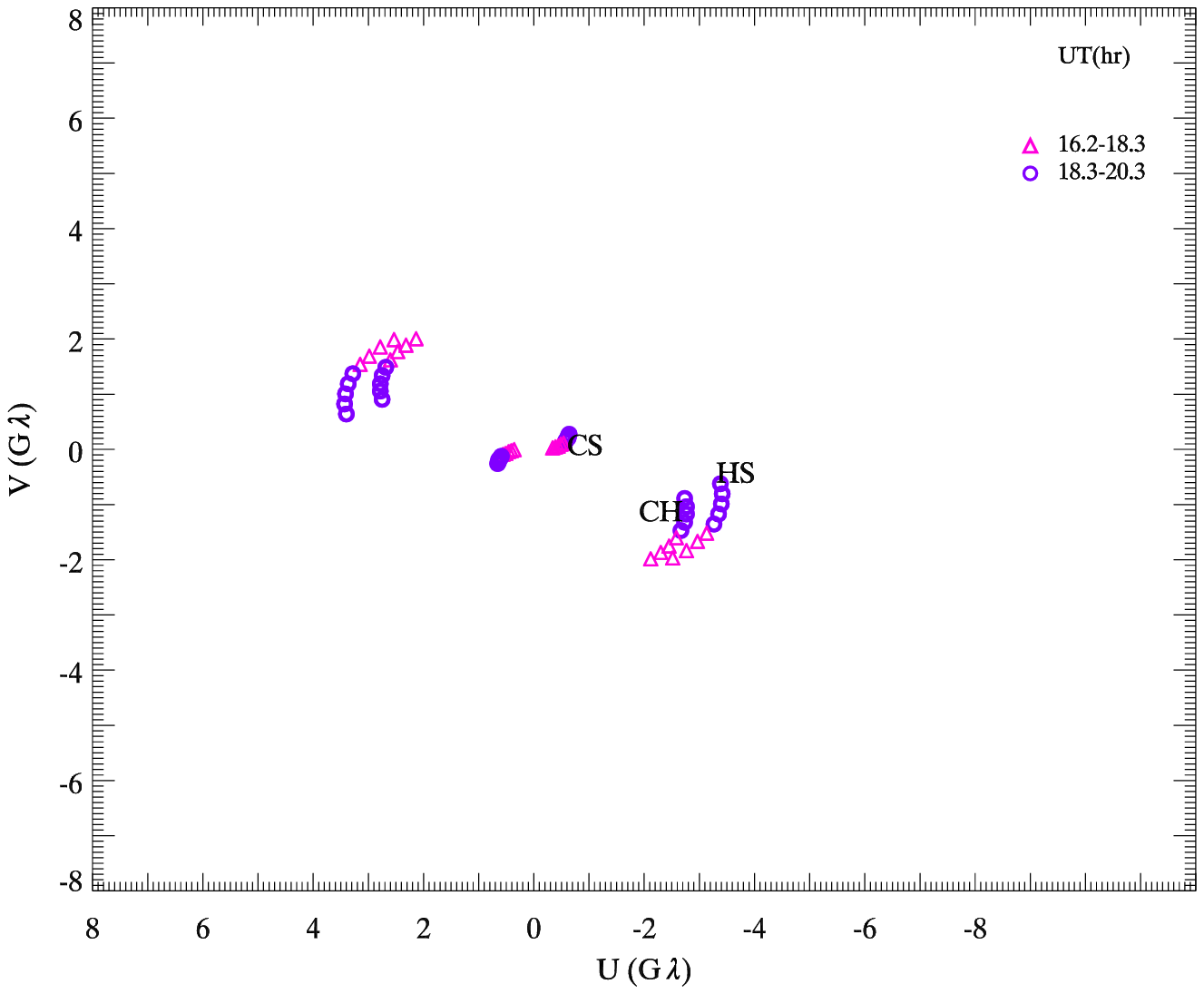} \\
\includegraphics[width=7.5cm]{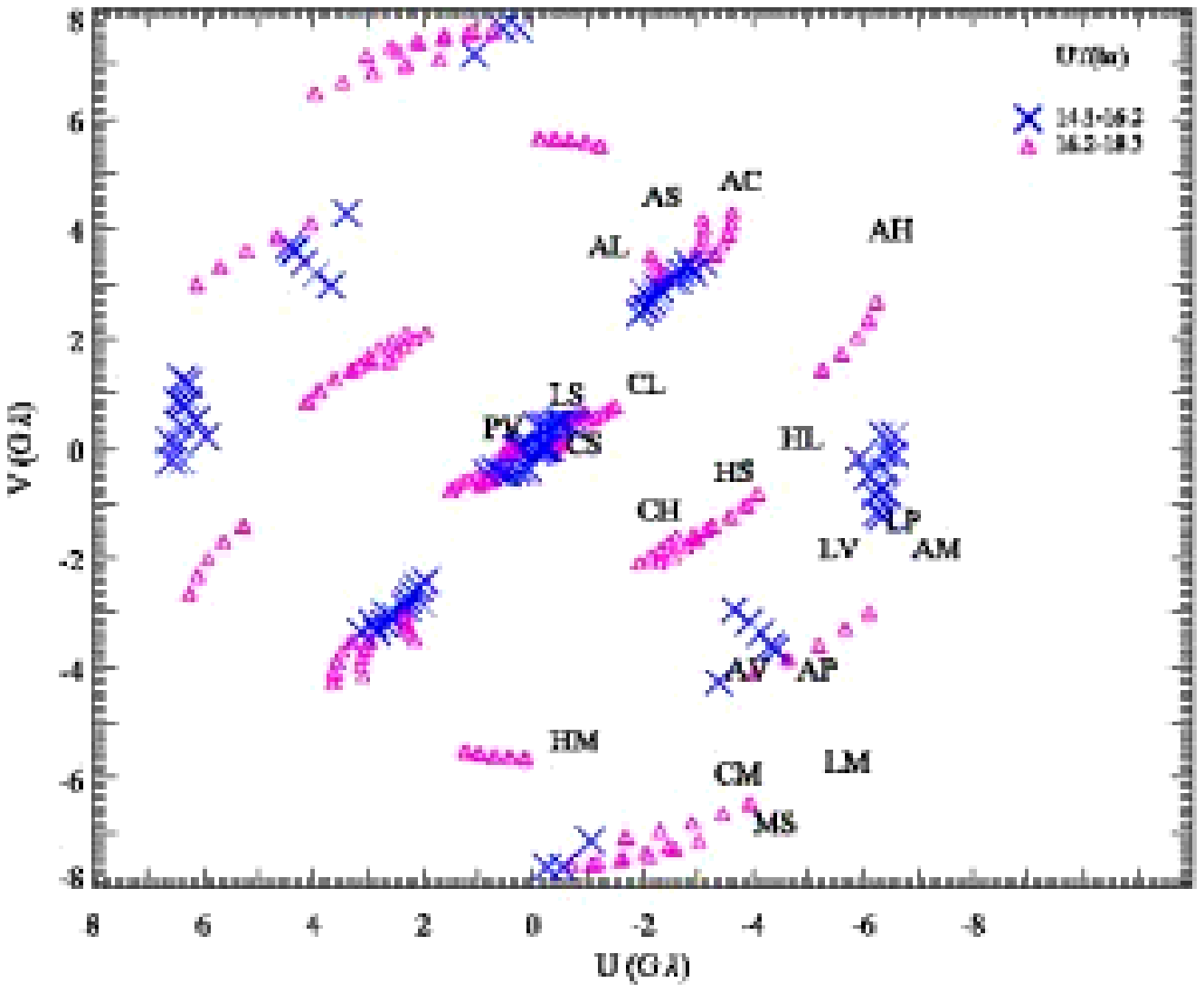}
\includegraphics[width=7.5cm]{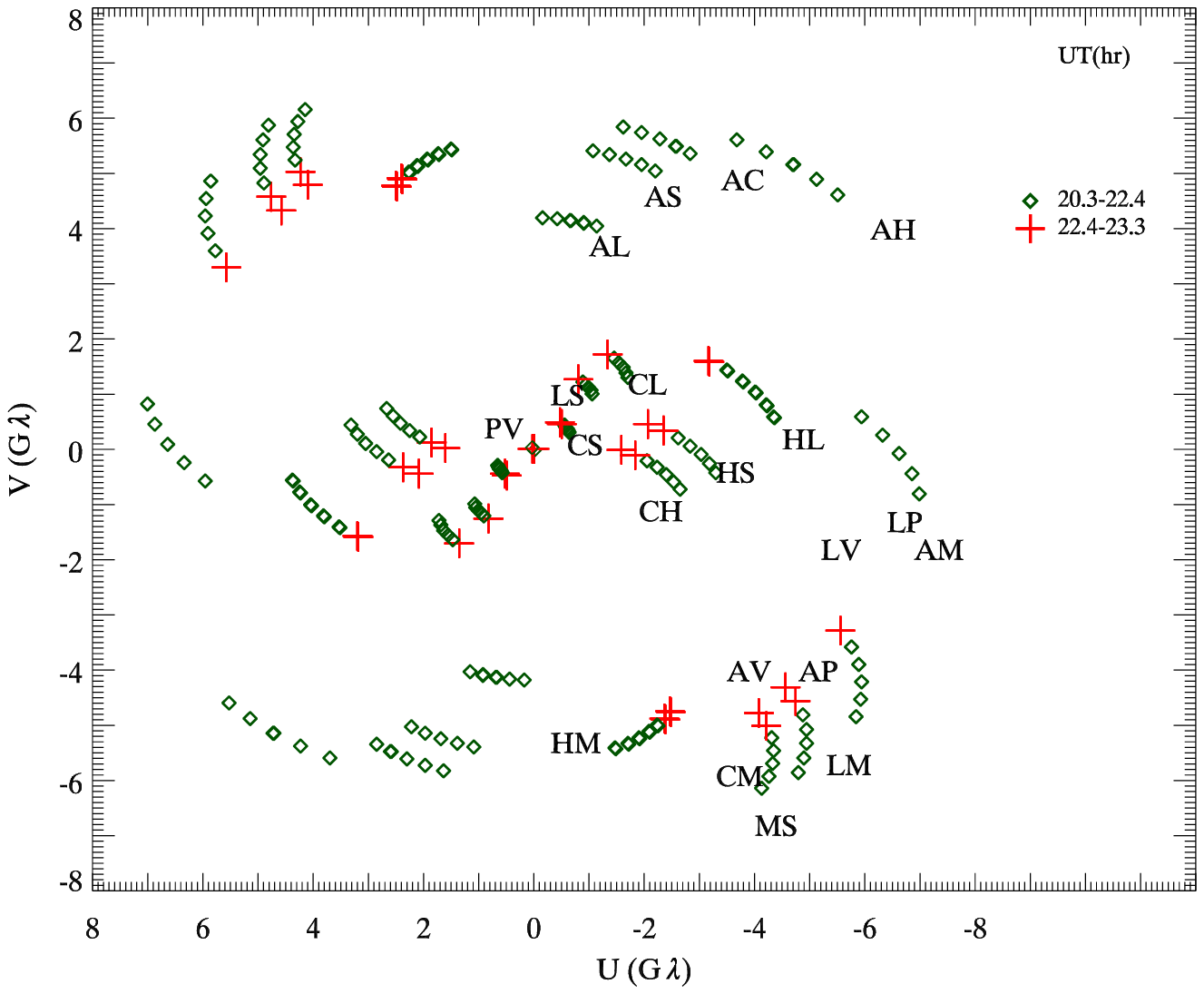}
\vspace{-5mm}\caption{ \textit{Top-left}: $(u,v)$ coverage produced
by eight potential stations at 1.3mm wavelength. H: Hawaii,
including JCMT and SMA, S: SMTO, C: CARMA, L: LMT, A: Chilean,
including ASTE and ALMA, P: PdBI, V: PV, M: MT-COOK. Tracks in
different parts of observational time are shown in different symbols
(square, cross, triangle, circle, rhombus, and plus), with the
interval of $\sim 2$hr. \textit{Top-right}: \textit{sub-coverage i}
during $\sim 16 - 20$hr (UT), produced by H, S, and C.
\textit{Bottom-left}: \textit{sub-coverage ii} during $\sim 14 -
18$hr (UT), produced by all the eight stations.
\textit{Bottom-right}: \textit{sub-coverage iii} during $\sim 20 -
23$hr (UT), produced by all the eight stations. \label{uv}}
\end{center}
\end{figure}

We consider the following eight potential mm-VLBI stations to
simulate the 1.3mm VLBI observations in this paper: Hawaii (H),
including JCMT and Submillimeter Array (SMA); SMTO (S); CARMA (C);
the Large Millimeter Telescope (LMT, L) on Sierra Negra, Mexico;
Chilean station (A), consisting of Atacama Submillimeter Telescope
Experiment (ASTE) and Atacama Large Millimeter Array (ALMA); the
IRAM Plateau de Bure (PdBI, P), France; the IRAM Pico Veleta, Spain
(PV, V); and a proposed station locating on Mount Cook in New
Zealand (MT-COOK, M), one of the candidate positions for telescope
establishment in the future. The $(u,v)$ coverage from these
stations is shown in the top-left panel of Fig.\ref{uv}. These
stations offer a $(u,v)$ coverage in almost all the direction and
various baseline lengths. However, it takes about 11 hours to
complete these tracks due to Earth rotation. Here we divide the
total observing time into six parts, with the interval of $\sim
2$hr. In Fig.\ref{uv}, they are shown in square(black), cross(blue), triangle(magenta),
circle(violet), rhombus(green), and plus(red) symbols, respectively. from the beginning
to the end. The $(u,v)$ coverage in the same or neighbour symbol, is
considered for a simultaneous observation.

We choose three typical sub-coverages from the total observation.
\textit{Sub-coverage i} is a sub-coverage during $\sim 16 - 20$hr
(UT), produced by the Hawaii (JCMT included), SMTO, and CARMA, shown
in triangle and circle in the top-right panel of Fig.\ref{uv}. This
sub-coverage includes the coverage obtained in the observations by
\citet{Doel08}. \textit{Sub-coverage ii} is during $\sim 14 - 18$hr
(UT) from all the stations considered, shown in cross and triangle
in the bottom-left panel in Fig.\ref{uv}. In this sub-coverage, a
wide range of baseline length of $\sim 0-5.5{\rm G}\lambda$ in
northwest-southeast direction is included. Especially, lengths of
$\sim 3-5{\rm G}\lambda$ are covered by baseline group AC/AL/AS. In
almost its perpendicular direction, i.e. the northeast-southwest
direction, comparable lengths are also covered by baselines CH, HS,
and HL. Simultaneous detections on both perpendicular directions are
very important to determine the geometry of the assumed accretion
flow, as shown in \citet{Huang07} and as follows.
\textit{Sub-coverage iii} is also from all the eight stations, but
during $\sim 20 - 23$hr (UT), shown in rhombus and plus in the
bottom-right panel. In this sub-coverage, a wide range of baseline
length of $\sim 0-7{\rm G}\lambda$ in near east-west direction is
covered by baselines CS, CH, HS, HL, and AM. The baselines AL, AS,
AC, AH, HM, CM, MS, and LM cover a wide range of directions apart
from the east-west, but only in lengths of $\gtrsim 4{\rm
G}\lambda$.

\section{Visibility Analysis And Parameter Estimation}
\label{vis}

\subsection{Black Hole Spin Estimation For the Fiducial Model}

\begin{figure}[h]
\vspace{-0mm}
\begin{center}
\includegraphics[width=7.5cm]{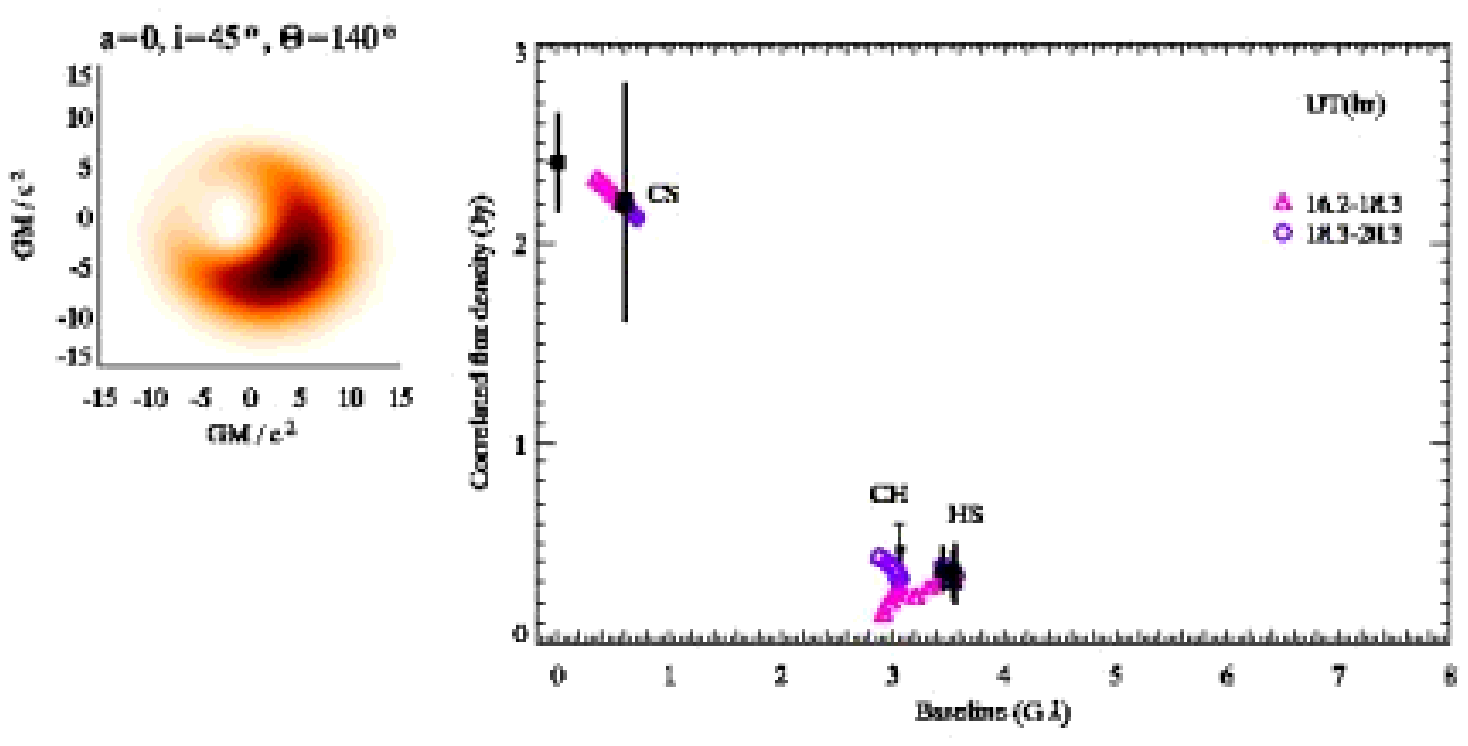} \\
\includegraphics[width=7.5cm]{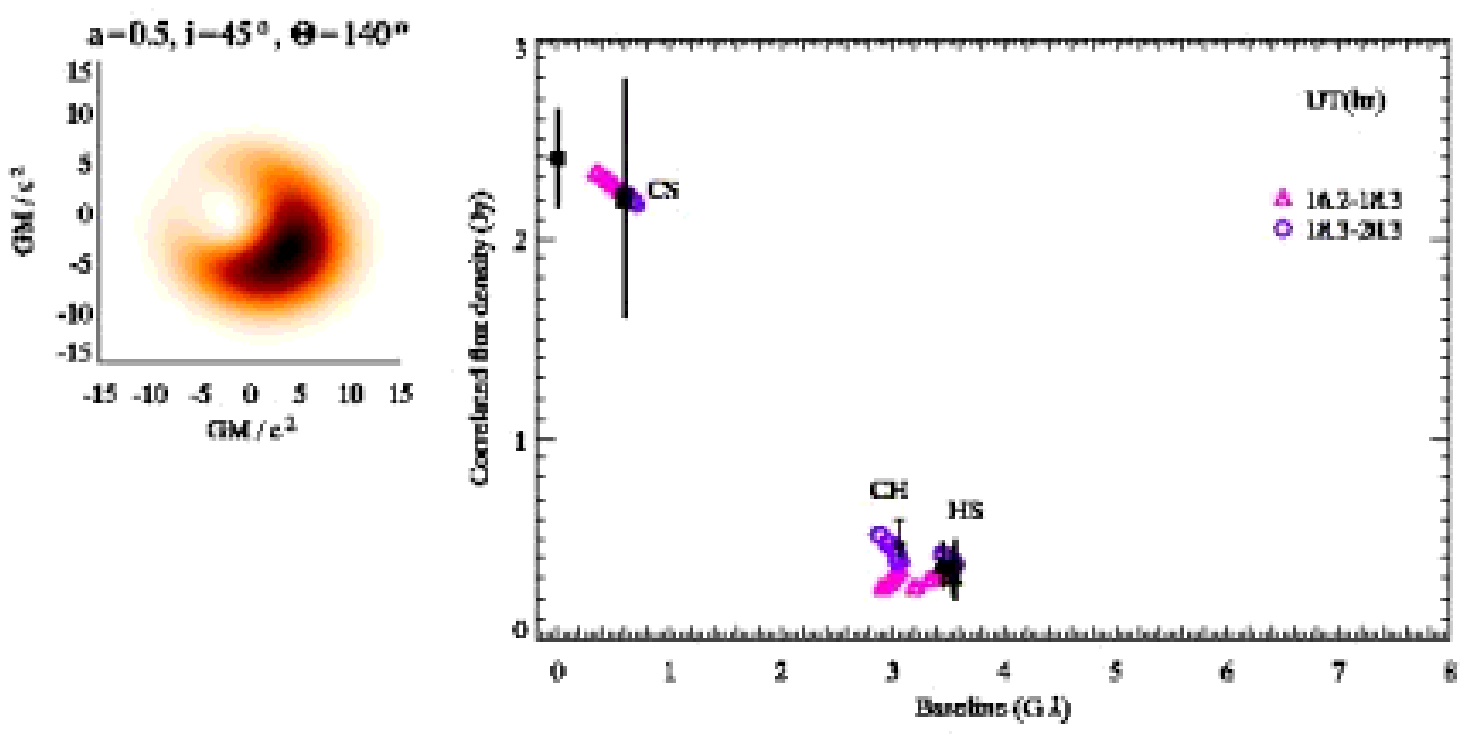} \\
\includegraphics[width=7.5cm]{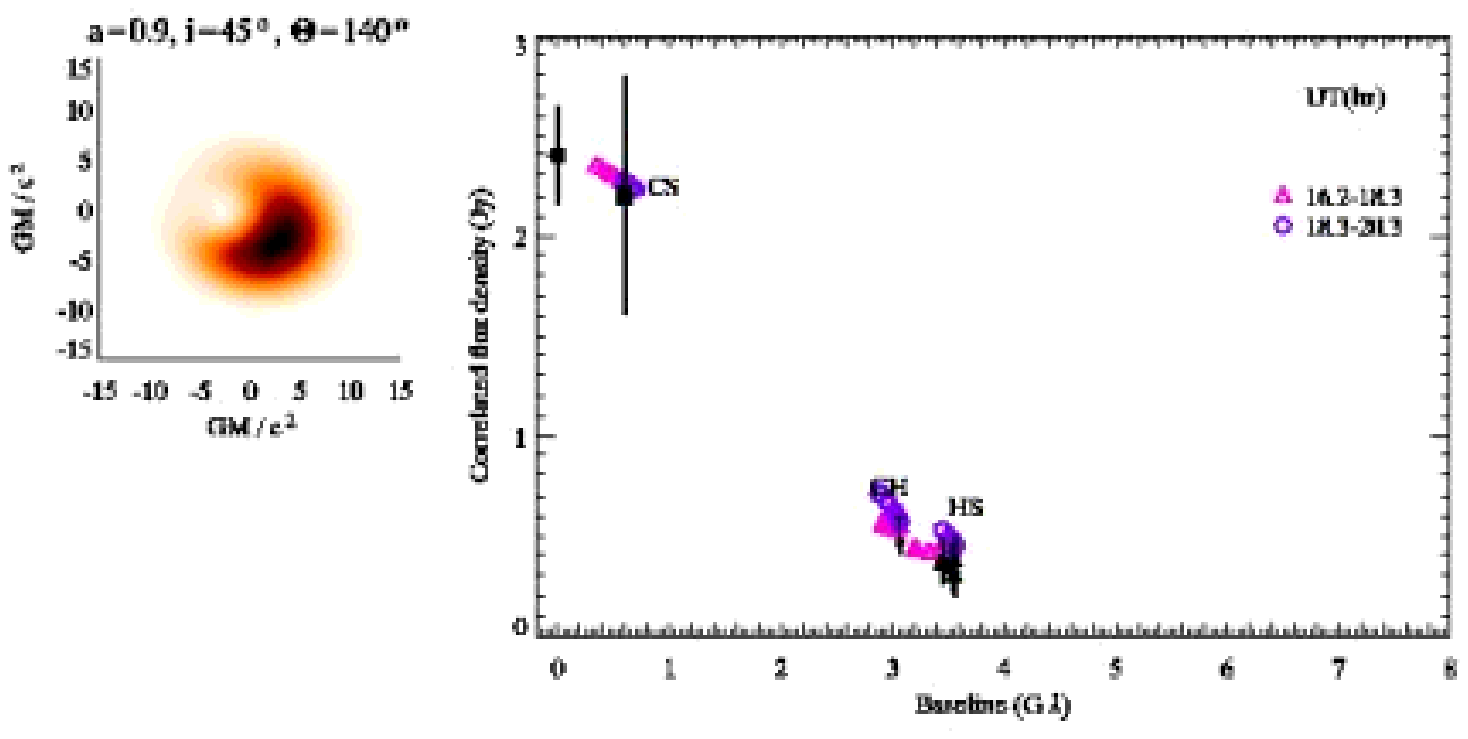}
\vspace{-5mm}\caption{
Images and visibilities predicted by the fiducial model with $a=0$, $0.5$, and $0.9$,
with the sampling of \textit{sub-coverage i}. The observational data are from
\citet{Doel08}.
\label{vispa}}
\end{center}
\end{figure}

For the fiducial model described in Sec.\ref{mod}, with inclination
angle $i=45^\circ$, position angle $\Theta=140^\circ$, but with
three different black hole spins, $a=0$, $0.5$, and $0.9$, we
convolve the images with the scattering ellipse. We adopt mean FWHMs
of the scattering screen of $1.39 \lambda^2$ for the major angular
size in milli-arc-second and $0.69 \lambda^2$ for the minor one, and
the orientation of $80^\circ$ for the major axis, which is derived
from size measurements in centimeter band \citep{Shen05}. We then
perform 2-dimensional Fourier transform with special $(u,v)$
coverage provided by those potential stations. With the sampling of
\textit{sub-coverage i}, the corresponding visibility profiles are
shown in Fig.\ref{vispa}. For each case, the scatter-broadened image
is shown in the left. Its black hole shadow structure is shown as a
white region at the center. The dark-red region beside the shadow
shows the accretion flow emission boosted by Doppler effect. The
predicted visibilities are shown in triangle and circle, with the
$(u,v)$ coverage in \citet{Doel08} included. As mentioned in
\citet{Fish09}, Sgr A* could be considered as observed in its
quiescent state since it exhibited the total flux density lower than
other measurements. These data can be compared with our predictions
within 4 hours, although they were measured on two consecutive days.
Their measurements of total flux density, correlated flux density on
baseline CS, and correlated flux density on baseline HS are marked
by filled circle, filled square, and filled triangles, respectively.
The upper limit on baseline CH is marked by a downward arrow.

The models with $a=0$ and $a=0.5$ provide good explanations to the
visibility measurements. Notice that the predicted visibilities in
triangle at baselines CH and HS, in similar directions, cannot be
fitted by a Gaussian profile. Clear structures of null point, or
valley point, are predicted at $\sim 3{\rm G}\lambda$, which are due
to the existence of the black hole shadow \citep{Miyo07,Huang07}. We
note that this structure may be absent in a lot of cases (see in the
following). In our fiducial model, however, the appropriate lengths
and directions make these two baselines, CH and HS, potential to
detect the important structure, if better sensitivity can be
achieved for CARMA.

When the black hole is assumed to be fast rotating, the predicted
correlated flux densities on baselines CH and HS also increase. This
is because that the emission region become more compact with large
spin. When the black hole spin increases to $0.9$, the predicted
visibilities on baseline CH nearly exceed the observed upper limit.
Furthermore, the structure of the valley point cannot be recognized.
The emission region becomes so compact and totally dominated by the
blue-shifted side. Therefore, for the fiducial model with
$i=45^\circ, \Theta=140^\circ$, determined by polarization data, the
1.3mm visibility measurements prefer a black hole spin parameter of
$a<0.9$.

\subsection{Orientation Estimation With 1.3mm VLBI}

The fiducial model with a relatively mildly rotating black hole can
reproduce recent 1.3mm visibility measurements. The orientation of
the accretion flow is constrained by polarization observations,
based on the assumption of a large-scale magnetic field in toroidal
configuration. However, this specific orientation is not conclusive
without direct imaging. In spite of the polarizations, the
measurements on baselines CH and HS can be explained by many
appropriate combinations of $a$, $i$, and $\Theta$, e.g.,
highly-inclined disk with $\Theta \sim 90^\circ - 150^\circ$ for
relatively mild spin, or nearly face-on disk with any position angle
and spin. Our favorite parameters are relatively mildly-rotating
black hole, mild inclination angle, and position angle toward
southeast. On the other hand, any model in opposite parameters,
i.e., nearly face-on, nearly egde-on, position angle toward
northeast, or extremely high spin, can be an extreme alternative to
our favorite model.

\begin{figure}[h]
\vspace{-0mm}
\begin{center}
\includegraphics[width=7.5cm]{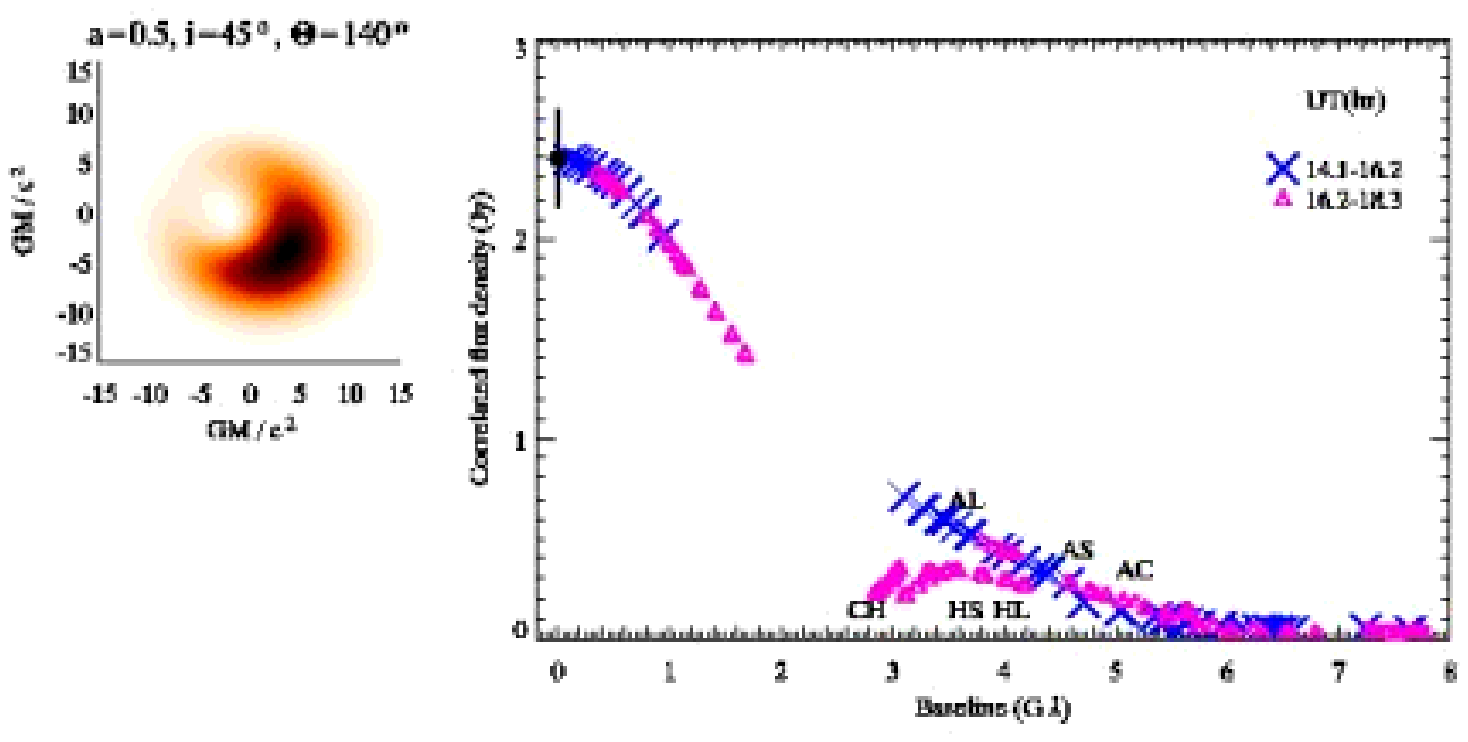}
\includegraphics[width=7.5cm]{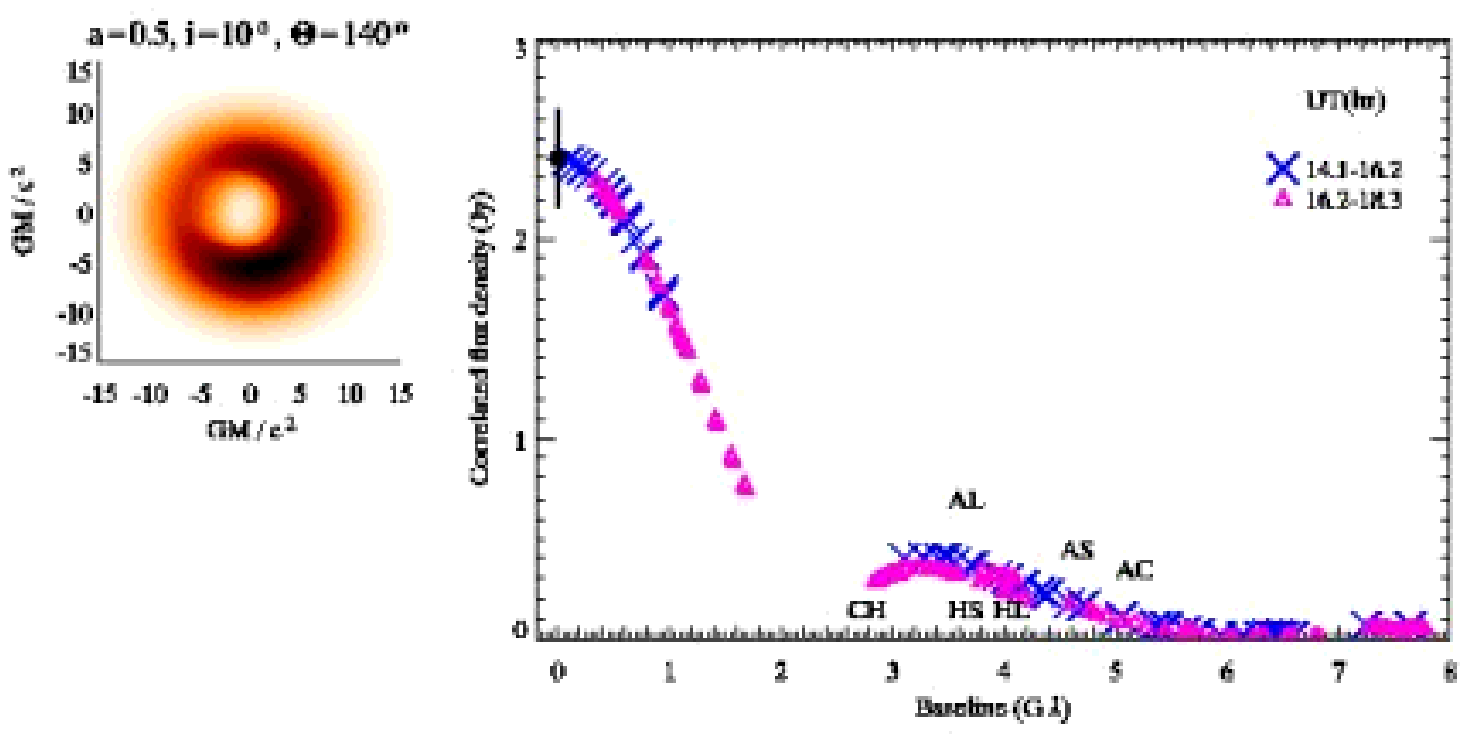} \\
\includegraphics[width=7.5cm]{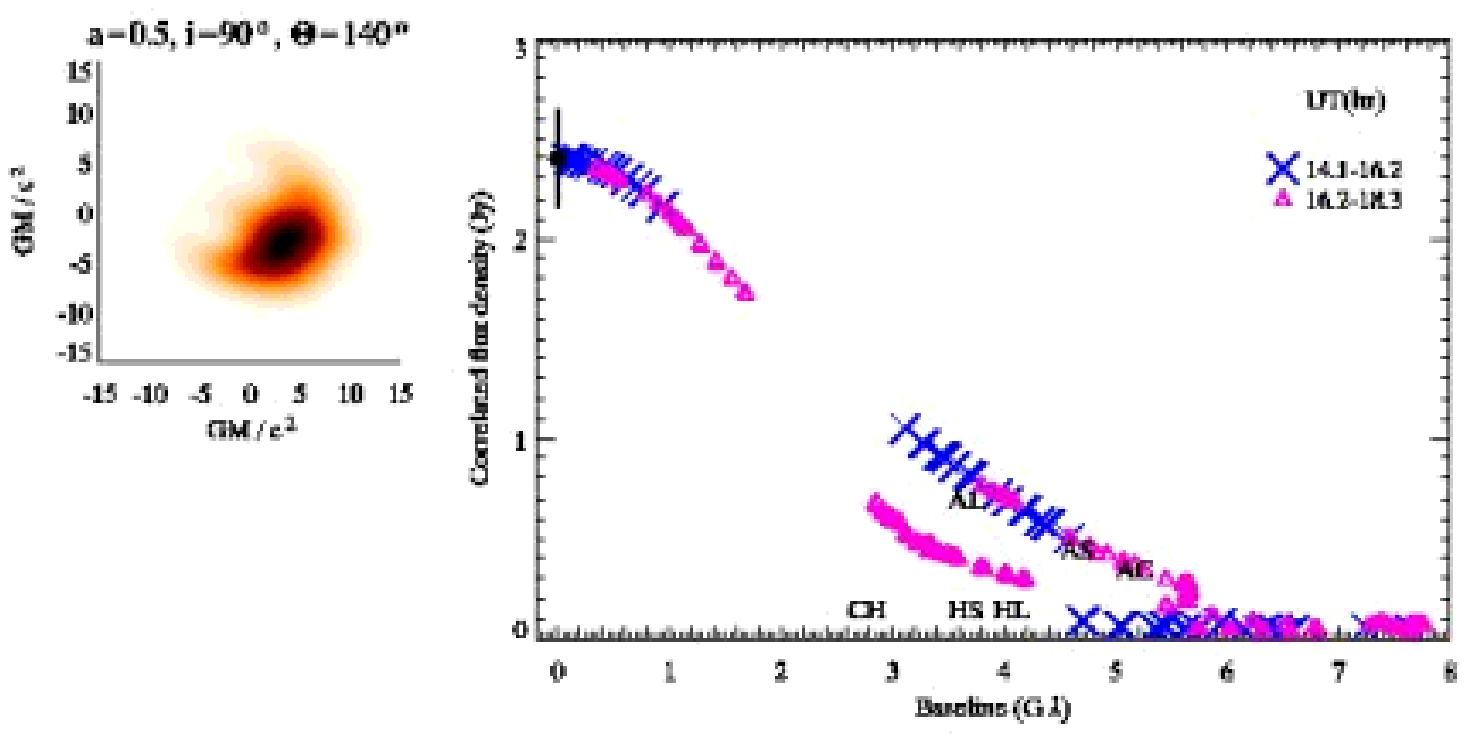}
\includegraphics[width=7.5cm]{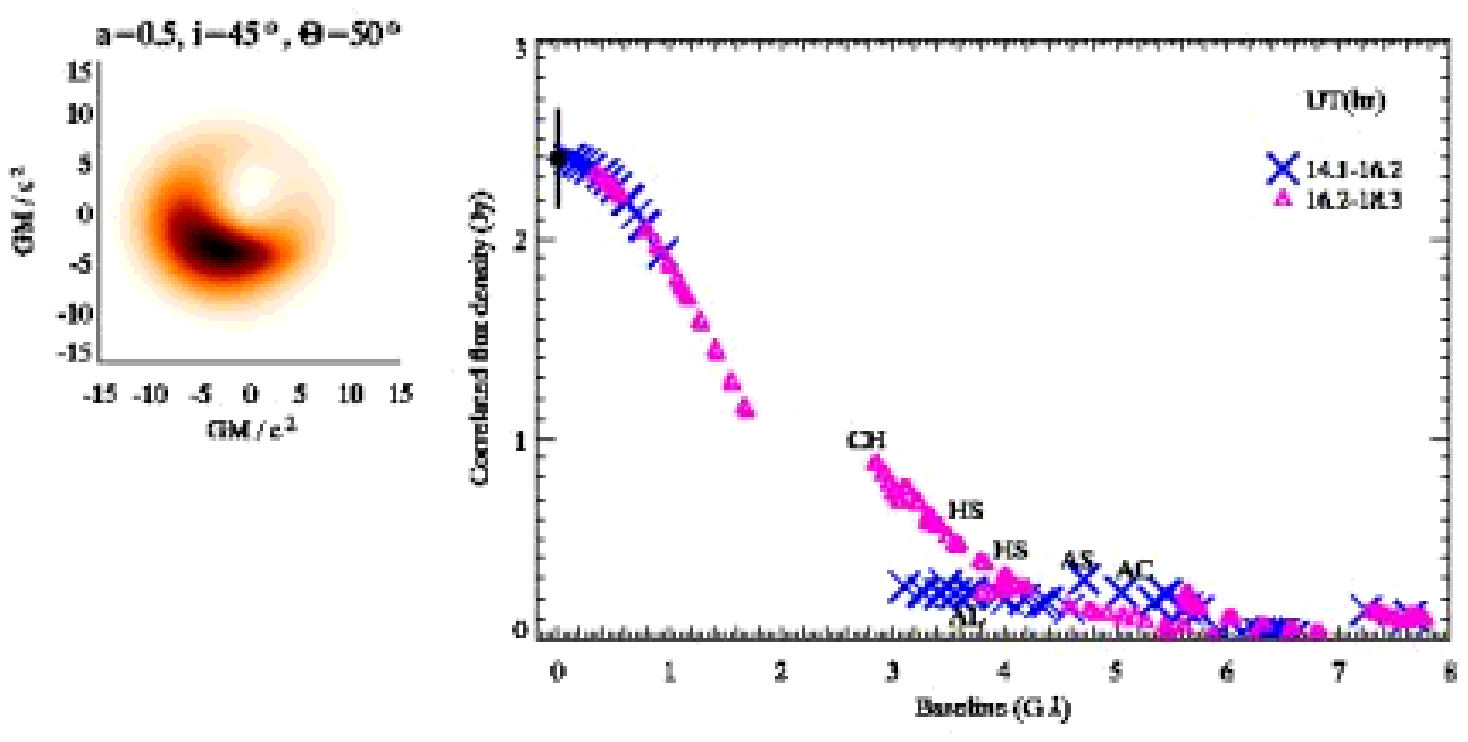}
\vspace{-5mm}\caption{
Images and visibilities predicted by four cases with $a=0.5$ but different orientation,
with the sampling of \textit{sub-coverage ii}. In these four cases, the visibilities yielded by baseline group AC/AL/AS can be higher than, lower than, or comparable to the visibilities yielded by baseline group CH/HL/HS, see in context for details.
\label{vispb1}}
\end{center}
\end{figure}

We focus on the disk orientation, i.e., $i$ and $\Theta$, first.
Choose one fiducial model with $a=0.5, i=45^\circ,
\Theta=140^\circ$, and set three alternative cases with $a=0.5,
i=10^\circ, \Theta=140^\circ$, $a=0.5, i=90^\circ,
\Theta=140^\circ$, and $a=0.5, i=45^\circ, \Theta=50^\circ$. We
sample the predicted visibilities with the \textit{sub-coverage ii}
mentioned in Sec.\ref{cov}, which covers comparable baseline lengths
in almost perpendicular directions by baseline groups CH/HL/HS and
AC/AL/AS. In particular, baseline AL can perform simultaneous
observations with baseline HL at $\sim 4 {\rm G}\lambda$, and
baselines AL and AS can perform quasi-simultaneous observations
(difference within 4 hours) with baselines CH and HS at $\sim 3-3.5
{\rm G}\lambda$. The corresponding visibilities are predicted in
Fig.\ref{vispb1}, shown in cross and triangle. It can be found that
the visibilities predicted by the two baseline groups make big
differences among all the possible cases. We can divide the
visibility results into three groups: Group I, the predicted
visibilities yielded by the two orthogonal baseline groups are
comparable, e.g., the case shown in the top-right panel; Group II,
the predicted visibilities yielded by AC/AL/AS are in higher
correlated flux density compared to those yielded by CH/HL/HS, e.g.,
cases in the left two panels; Group III, opposite to Group II, e.g.,
the case in the bottom-right panel. These observations are capable
of determining the basic geometry of the intrinsic emission region
of Sgr A*. Detailed discussions are as follows.

\begin{figure}[h]
\vspace{-0mm}
\begin{center}
\includegraphics[width=7.5cm]{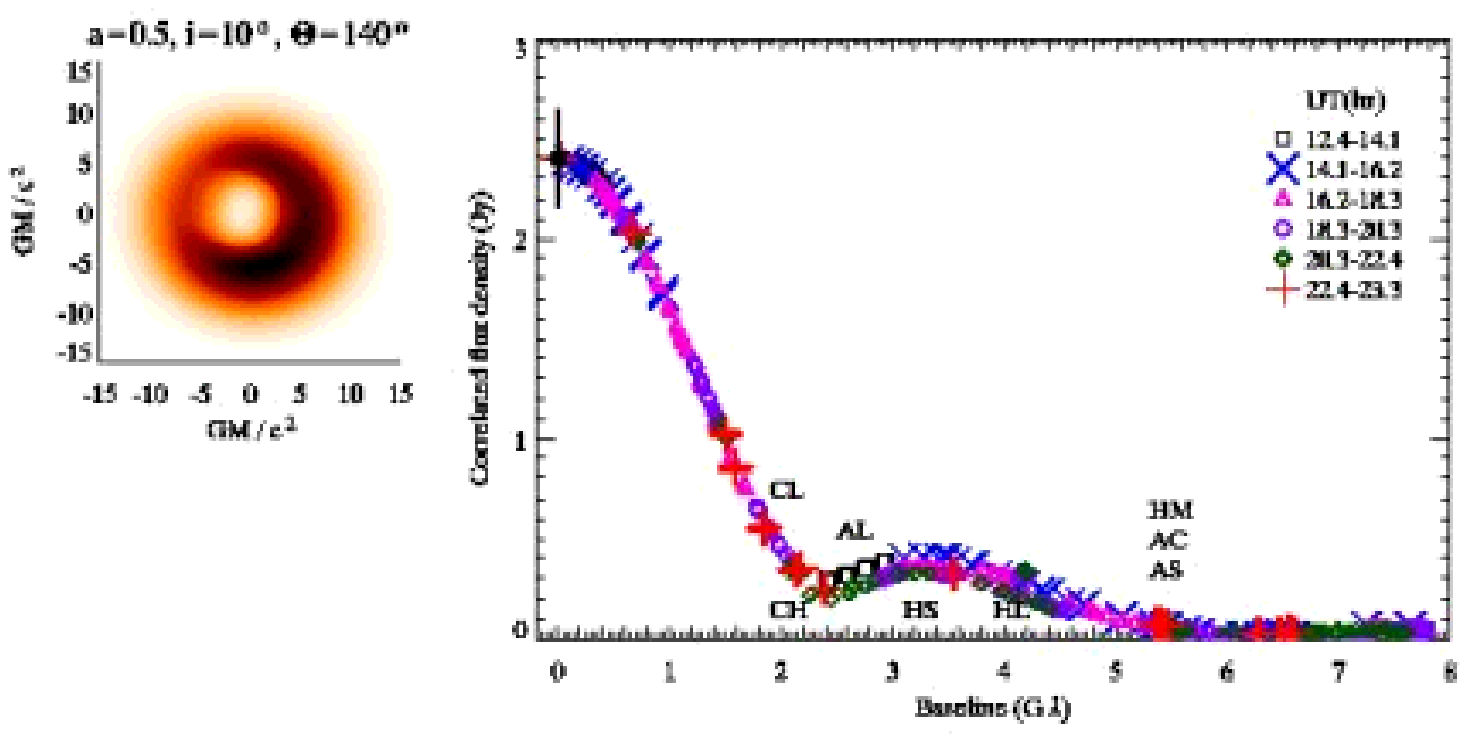}
\includegraphics[width=7.5cm]{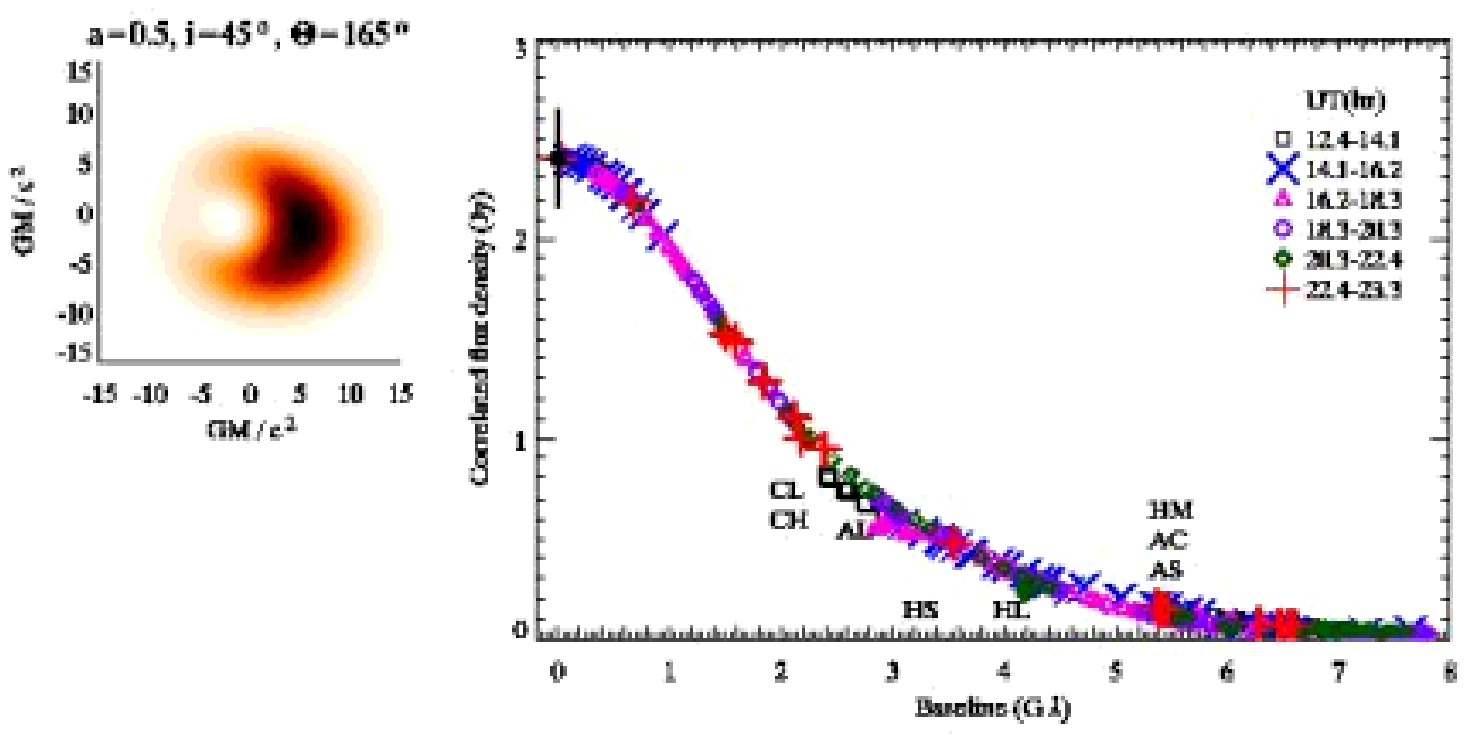}
\vspace{-5mm}\caption{ Images and visibilities predicted by two
cases with $a=0.5$ but different orientation combinations, with the
sampling of the total $(u,v)$ coverage. In both the cases, the
visibilities yielded by baseline group AC/AL/AS are comparable to
the visibilities yielded by baseline group CH/HL/HS. \label{vispb}}
\end{center}
\end{figure}

If the results in Group I are obtained, the emission region is
almost symmetrical to the east-west direction. It is probably the
nearly face-on case, similar to that shown in the top-right panel of
Fig.\ref{vispb1}. We show its visibilities by the total $(u,v)$
coverage in the left panel of Fig.\ref{vispb}. In such a case, the
visibilities show a similar profile, with a clear valley along all
directions. Baselines with appropriate length, $\sim 2-3 {\rm
G}\lambda$ in this case, in any direction, such as CH, HS, CL, AL,
have the potential to detect such structure. However, there are
exceptions. Cases with higher inclination angle can be included in
Group I, if the position angle is close to north or south. We show
an example with $a=0.5, i=45^\circ, \Theta=165^\circ$ in the right
panel of Fig.\ref{vispb}. In this case, the visibilities again show
a similar profile along all directions, though for all the baselines
we consider here the valley is missing. However, these profiles
significantly deviate from Gaussian profiles.

\begin{figure}[h]
\vspace{-0mm}
\begin{center}
\includegraphics[width=7.5cm]{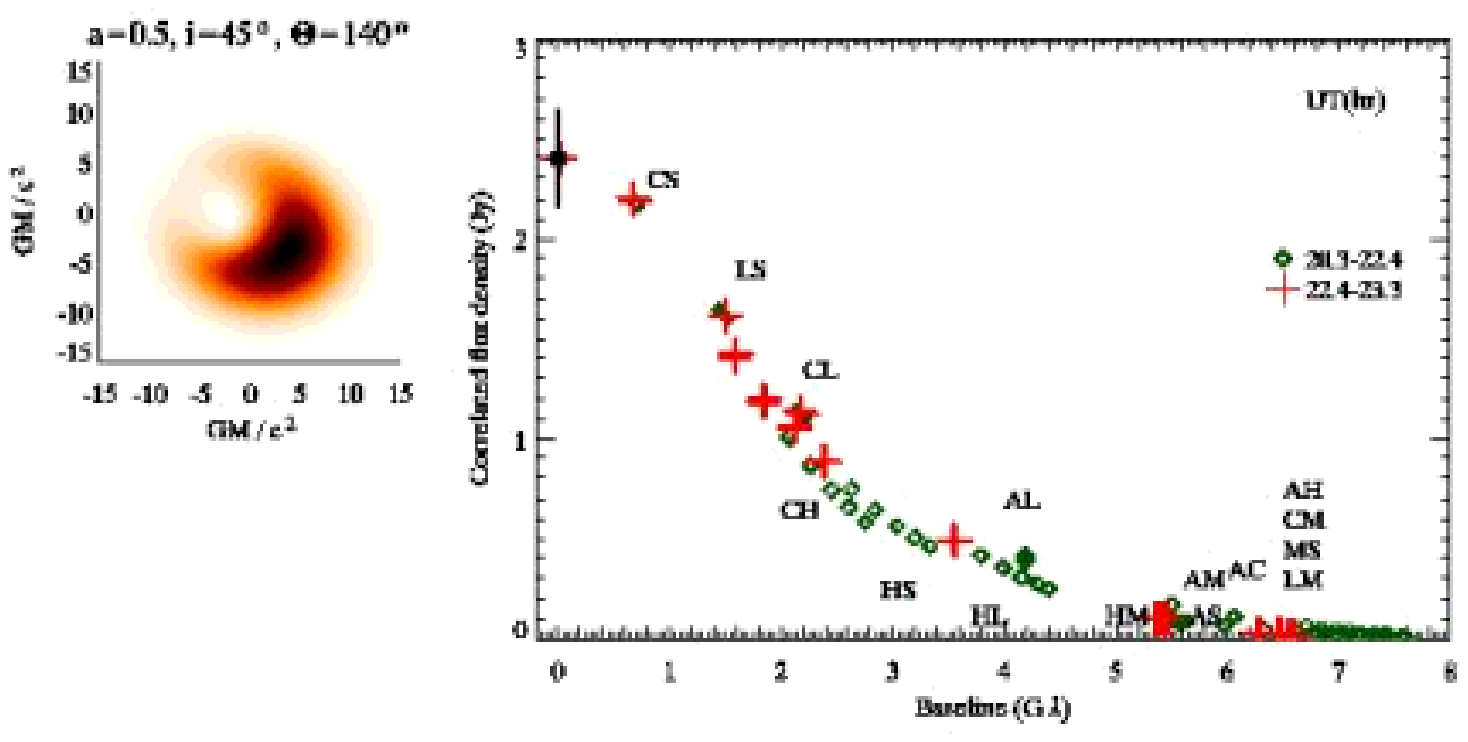}
\includegraphics[width=7.5cm]{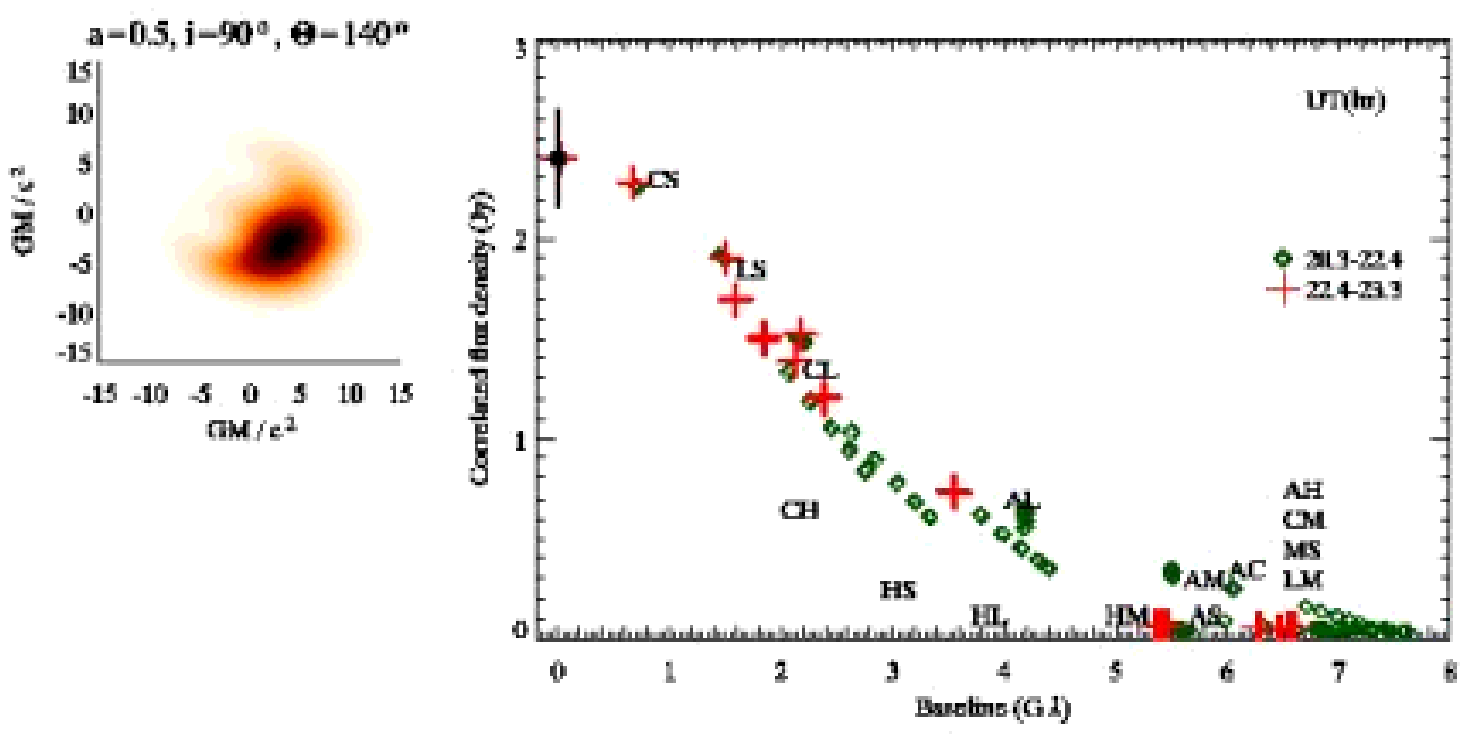} \\
\includegraphics[width=7.5cm]{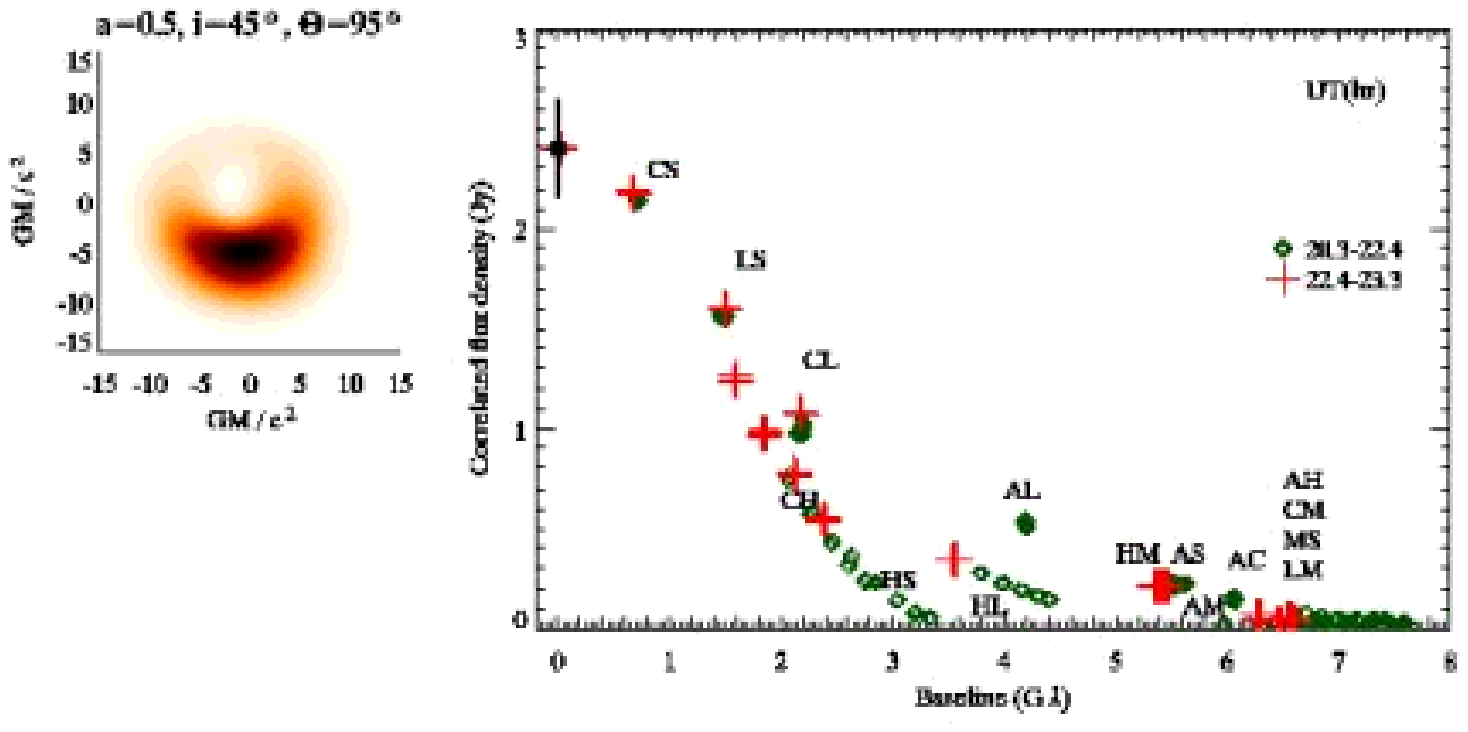}
\includegraphics[width=7.5cm]{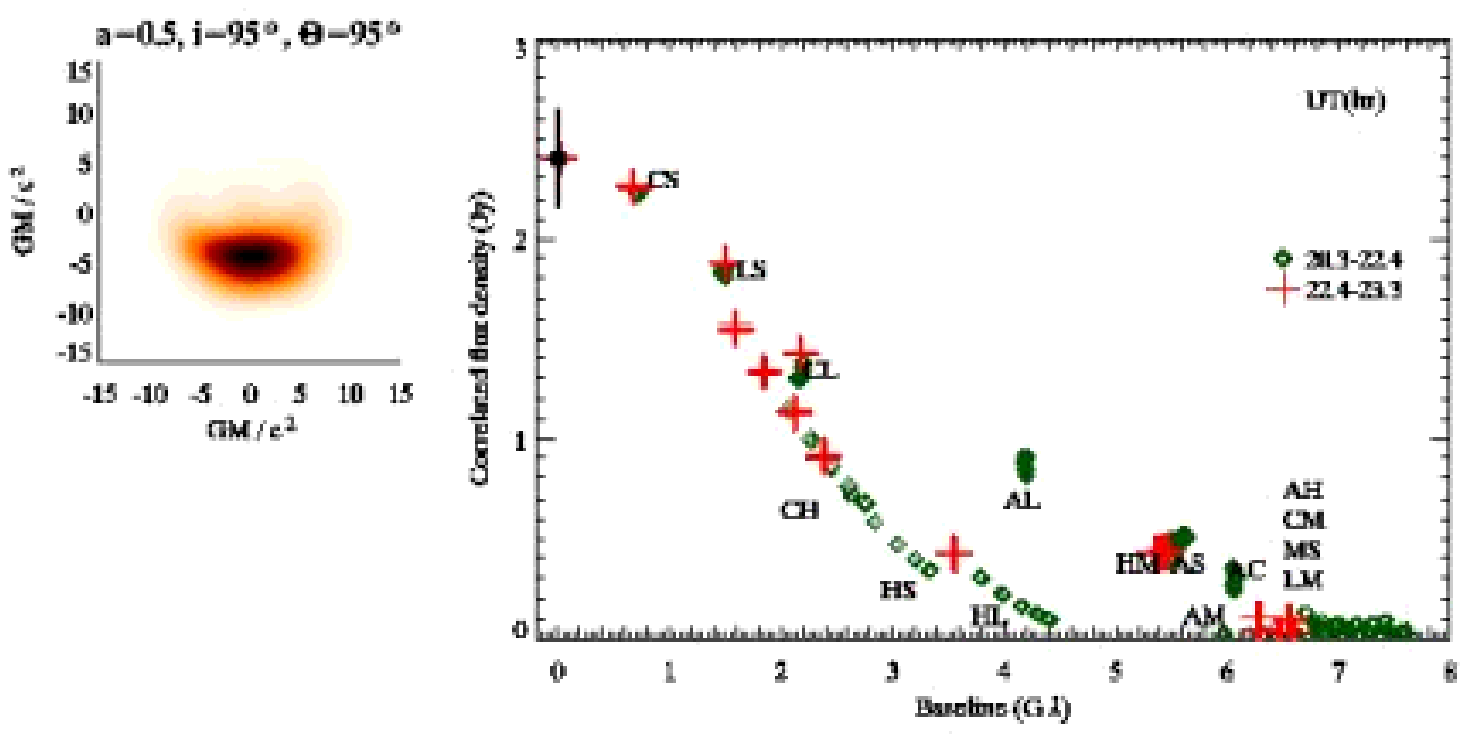}
\vspace{-5mm}\caption{ Images and visibilities predicted by four
cases with $a=0.5$ but different orientation combinations, with the
sampling of \textit{sub-coverage iii}. In these four cases, the
visibilities yielded by baseline group AM/CH/HL/HS can be higher
than, lower than, or comparable to the visibilities yielded by
baseline group AC/AH/AL/AS/CL/LS, see in context for details.
\label{vispb4}}
\end{center}
\end{figure}

If the results in Group II are obtained, the nearly face-on cases
are excluded, and the position angle is estimated ranging from $\sim
80^\circ$ to $\sim 150^\circ$, or from $\sim -30^\circ$ to $\sim
-100^\circ$. The structure of valley point may be detected on
baselines CH or HS at $\sim 2.5-3.5 {\rm G}\lambda$, except the
highly-inclined cases, i.e., $i \gtrsim 60^\circ$. In general, for a
given inclination angle, the differences in correlated flux density
between AC/AL/AS and CH/HL/HS are predicted to reach maxima at
$\Theta \sim 120^\circ$. For a given position angle, such
differences are predicted to be larger with a higher inclination
angle. VLBI observations reported in \citet{Doel08} are insufficient
to produce a unique estimate of the disk orientation in this case.
However, Earth-aperture synthesis with additional VLBI stations may
be able to do so. \textit{Sub-coverage iii} is complementary to 
\textit{sub-coverage ii} because of the different directions it covers. 
We sample the predicted visibilities with the new sub-coverage for the two cases
in left panels of Fig.\ref{vispb1}. We then make the same sampling
for these two cases by changing their position angles into
$95^\circ$. Predicted visibilities of all these four cases are shown
in Fig.\ref{vispb4}.

Two baseline groups are considered, AM/CH/HL/HS which provides
coverage on near east-west directions, and AC/AH/AL/AS/CL/LS which
provides coverage from northeast to northwest. These two baseline
groups can perform simultaneous observations with comparable
baseline lengths but significantly different directions. In
particular, baselines LS and CL track $\sim 45^\circ$ from baseline
CH $\sim 1.5-2.5 {\rm G}\lambda$. Baseline AL tracks nearly
perpendicular to baseline HL at $\sim 4 {\rm G}\lambda$. And AM and
AS track nearly perpendicular to baseline AM at $\sim 5-5.5 {\rm
G}\lambda$. With a position angle close to $\sim 150^\circ$ (or
$\sim -30^\circ$), e.g., the two cases with $\Theta=140^\circ$, the
visibilities yielded by the two baseline groups are predicted to be
comparable at $\lesssim 5 {\rm G}\lambda$, while in larger
difference with higher inclination angle at $> 5 {\rm G}\lambda$,
i.e. between AC, AS, HM, and AM. If the position angle decreases
from $\sim 150^\circ$ to $\sim 80^\circ$ (or from $\sim -30^\circ$
to $\sim -100^\circ$), e.g., the two cases with $\Theta=95^\circ$,
the visibilities yielded by group AM/CH/HL/HS are predicted to
become much lower than those yielded by group AC/AH/AL/AS/CL/LS.
With higher inclination angle, these differences become larger, in
particular, between AL and HL at $\lesssim 4 {\rm G}\lambda$, and
between HM (or AS) and AM at $\lesssim 5-5.5 {\rm G}\lambda$. In
addition, the valley structure, or deviation from Gaussian may
appear at baselines HL and HS at $\lesssim 3-4 {\rm G}\lambda$.

With the two sub-coverages discussed above, we may make an
estimation with an error bar as small as $\sim 20^\circ$ of the
position angle for an accretion flow of Group II. Furthermore, an
edge-on case may be easily recognized by visibilities at $> 5 {\rm
G}\lambda$.

Cases in Group III are not preferred by the recent 1.3mm measurements.
Those measurements have strictly constrained these cases
with $a \lesssim 0.5$, $20^\circ \lesssim i \lesssim 45^\circ$,
and $\Theta \sim 0^\circ$, so that we won't discuss more in this paper.
Of course, we can further use \textit{sub-coverage iii} for confirmation.

\subsection{Black Hole Spin Estimation With 1.3 mm VLBI}

Small values of black hole spin parameter are preferred by recent
measurements. However, this is not conclusive because it is based on
our fiducial model, with assumptions of specific electrons heating
mechanism and magnetic field structure. In \citet{Yuan09}, we adopted
the general radiatively inefficient accretion flow based on the work of
\citet{Manm00} and \citet{YQN03}, and found the recent measurements prefer extremely
high values of black hole spin, in stark contrast to the result
in this paper. Therefore, the spin parameter inferred from observational
visibilities depends significantly upon the adopted model.

\begin{figure}[h]
\vspace{-0mm}
\begin{center}
\includegraphics[width=7.5cm]{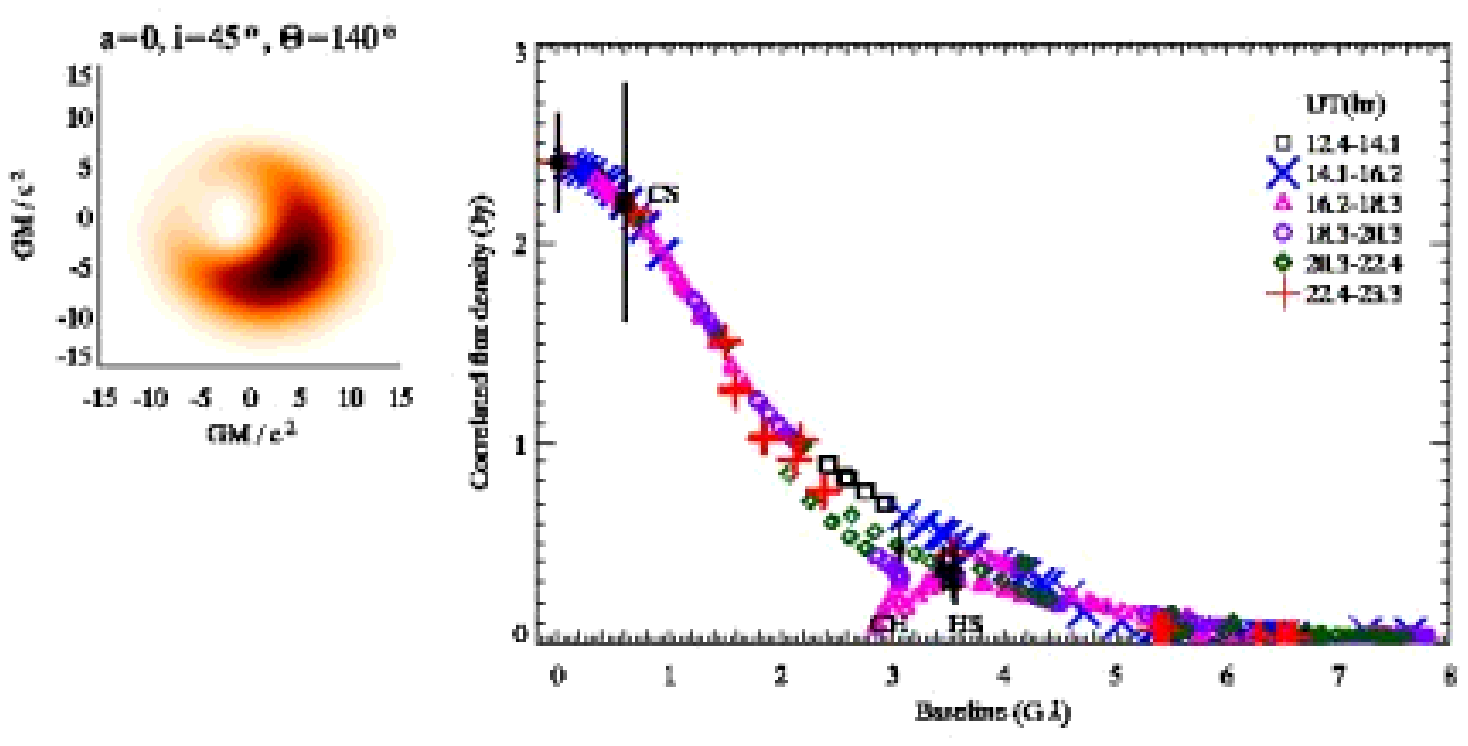}
\includegraphics[width=7.5cm]{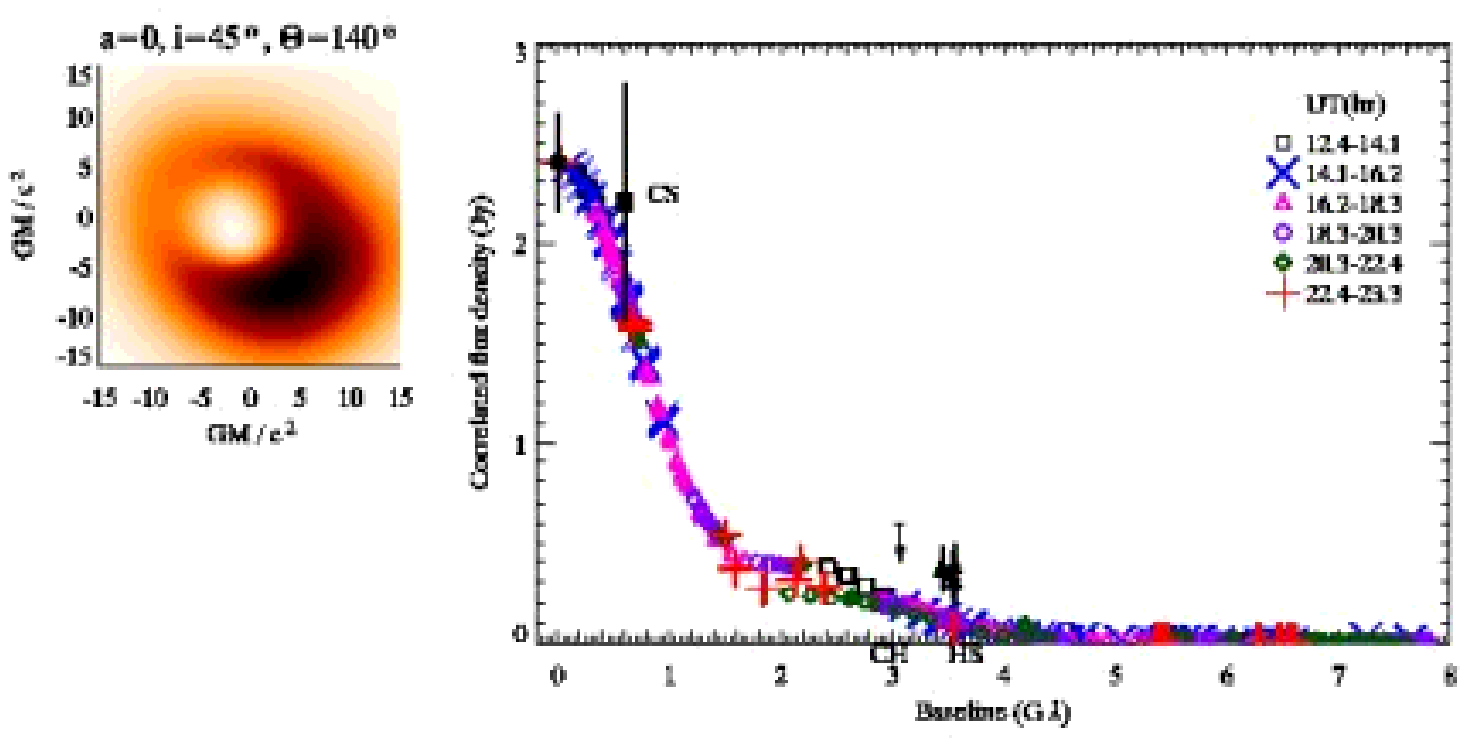} \\
\includegraphics[width=7.5cm]{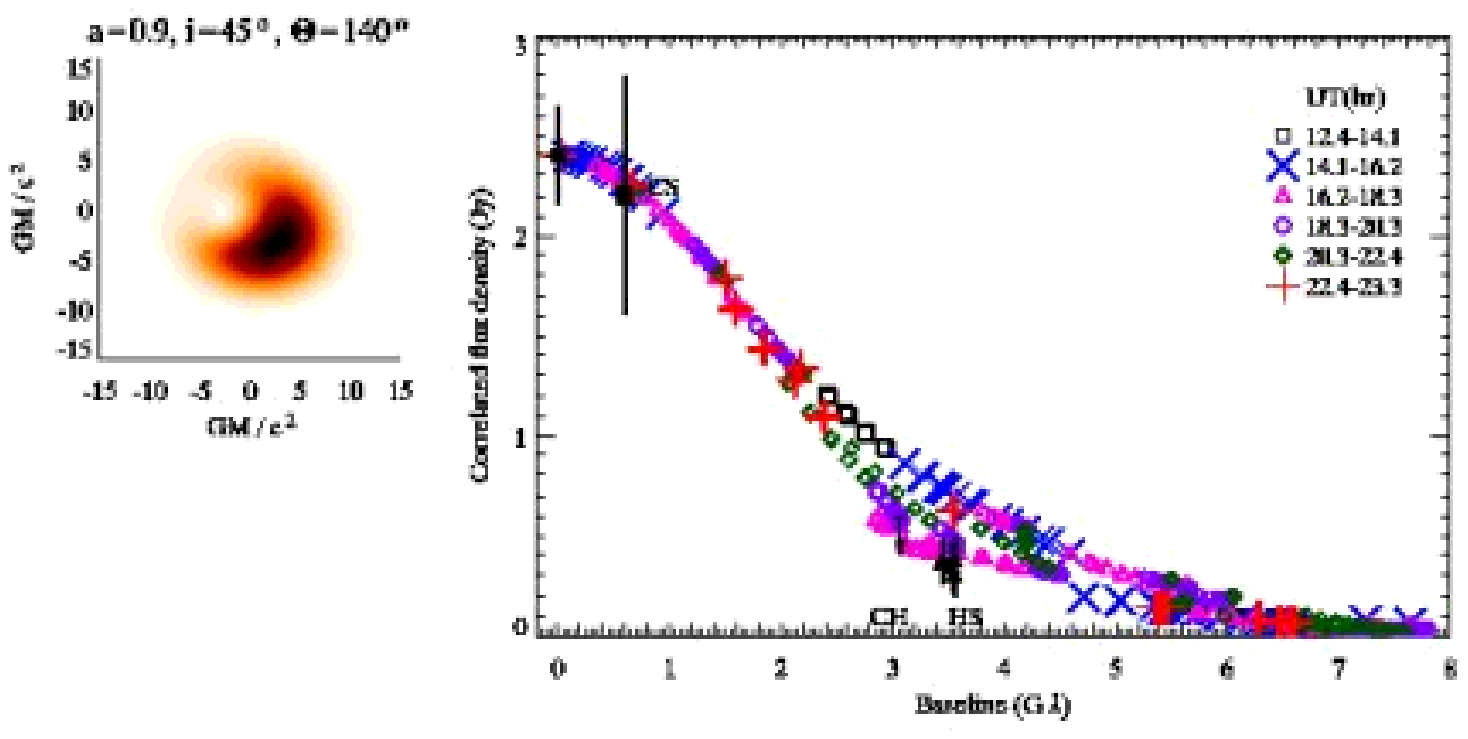}
\includegraphics[width=7.5cm]{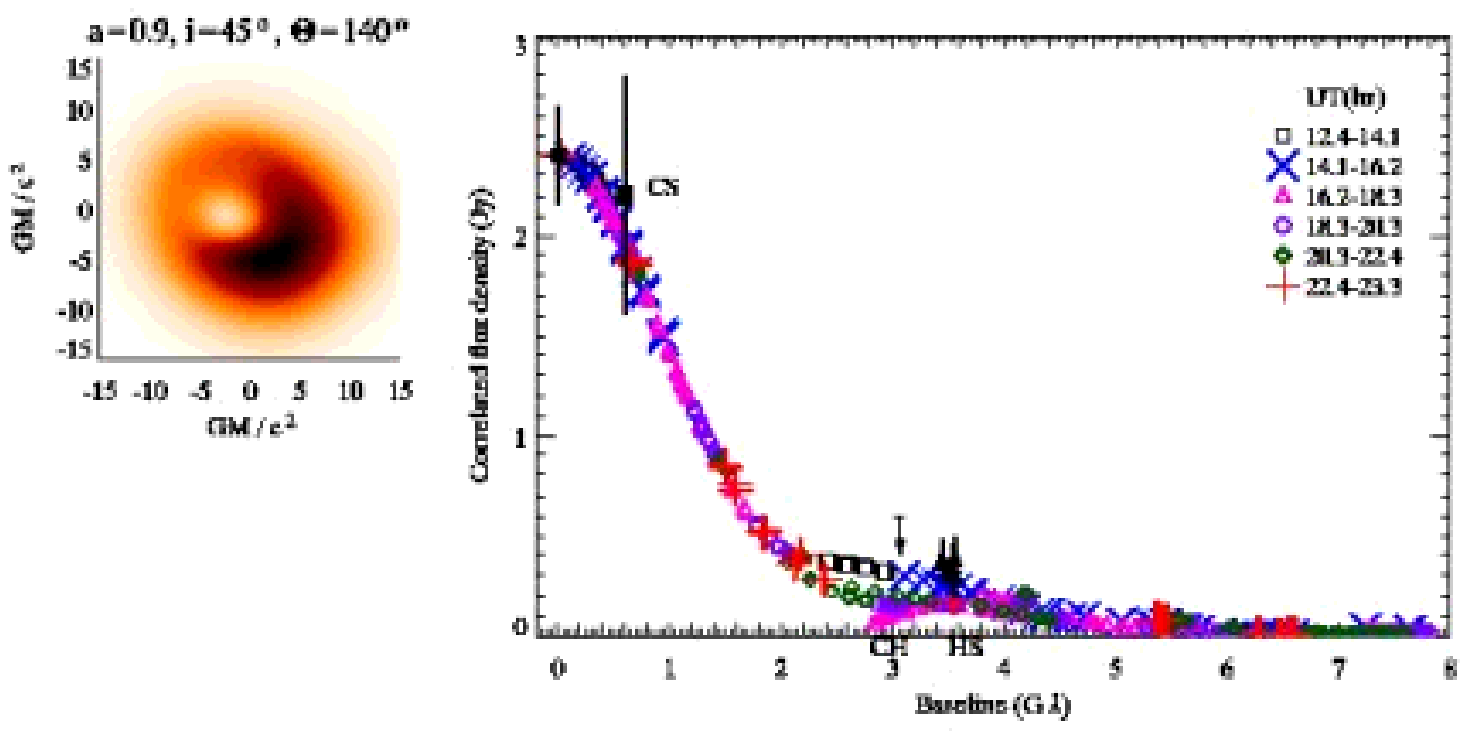}
\vspace{-5mm}\caption{ \textit{Left}: Images and visibilities
predicted by the fiducial model adopted in this paper (Model A) with
$a=0$ and $0.9$, with the sampling of the total $(u,v)$ coverage.
\textit{Right}: Images and visibilities predicted by radiatively
inefficient accretion flow model adopted in \citet{Yuan09} (Model
B), with the same spin, orientation, and $(u,v)$ sampling as the
left ones. \label{vispbspin}}
\end{center}
\end{figure}

Given a specific model, the emission region becomes more compact
with the increase of the black hole spin, so that the structure of
the visibility valley, if present, moves towards longer baselines.
What can be concluded from the visibility measurements is the
basic geometry, e.g., compactness, or asymmetry, of the emission
region. However, determining the black hole spin from this is
generally model-dependent. We show the images and predicted
visibilities of two cases for the fiducial model adopted in this
paper (Model A) in the left column in Fig.\ref{vispbspin}. For
comparison, two cases for the model adopted in \citet{Yuan09}
(Model B) with the same parameters of spin and orientation are
shown in the right column. It is found that the emission region of
Model A with a non-rotating black hole is even more compact than
that of Model B with a fast-rotating black hole, e.g., $a=0.9$.
Notice the predicted visibilies on baseline CH and HS, in circle
and triangle at $\lesssim 3-3.5{\rm G}\lambda$. and the data of
recent measurements. Fast-rotating black hole is excluded by Model
A, but preferred by Model B.

Such a difference is mainly caused by the different electron heating
mechanism used. In Model A, the heating mechanism is dominated by
turbulence in plasma wave. This heating mechanism is so effective
that the electron temperature exceeds $10^{11}$K in the innermost
region. The angular velocity of the accretion flow is close to the
Keplerian velocity, $\Omega \gtrsim 0.9 \Omega_K^+$, outside the
sonic point. The strong Doppler boosting in the blue-shifted side
makes the emission region more compact. In Model B, the electrons
are heated by a small fraction of the viscous heating. The electron
temperature keeps lower than $10^{11}$K even near to event horizon,
except in cases with extremely fast-rotating black hole. Compared to
Model A, it shows a flat profile in the innermost region, which
extends the emission region with the same spin. Moreover, the
angular velocity of the accretion flow is much lower than the
Keplerian velocity, $\Omega \lesssim 0.6 \Omega_K^+$. The much
weaker Doppler effect makes the emission region more extended.

In order to make a good estimation of the black hole spin
parameter, we should take advantage of the properties of the
space-time geometry that are independent of configuration of the
emission region. For example, the shape, size of the black hole
shadow, and its horizontal shift from the rotation axis of the
black hole \citep{Taka04} are the same for different models with
the same spin parameter, as if the emission region is assumed to
extend to the event horizon. Therefore, high-fidelity imaging via
VLBI at mm/sub-mm wavelengths is critical for this. Detecting the
innermost stable circular orbit of hot-spot structure, if exits,
during flaring activity is another way to constrain the spin
parameter directly \citep{BL05, BL06, Doel09}.

\section{Discussion}
\label{dis}

We consider eight potential stations to simulate the future 1.3mm
VLBI observations for Sgr A* to test the specific accretion flow
model with plasma wave heating mechanism. These stations provide a
highest angular resolution of $\sim 30 \mu$as, which is comparable
to the apparent angular size of the predicted black hole shadow
structure. For our fiducial model with $i \sim 45^\circ$ and $\Theta
\sim 140^\circ$, the recent 1.3mm VLBI measurements constrain its
black hole spin parameter as $a \lesssim 0.9$. More detailed
visibility analysis suggests that future VLBI observations can
improve these estimates substantially.

\subsection{Comparisons To Earlier Work}

As mentioned in the previous section, the black hole parameter estimation
is dependent on the specific model adopted to fit the observations.
This is born out by other efforts to fit the 1.3mm-VLBI results.

\citet{Brod09} followed the general RIAF model but adopted a Keplerian
velocity distribution. They preferred small values of black hole spin.
This is different from the prediction of extremely large spin from our
Model B, which may be caused by different treatments of
velocity field and gravitational effect, i.e. \citet{YQN03} used Pseudo-Newtonian
potential instead of Kerr metric.
Interestingly, their likeliest values of black hole parameter,
$a=0, i=40^\circ, \Theta=-23^\circ (157^\circ)$, are consistent with those
preferred in our Model A. The Keplerian, or near Keplerian, velocity field
generically makes the emission region more compact, preferring small spin parameter.

Their most probable combination is with values
$a=0, i=90^\circ, \Theta=-18^\circ (162^\circ)$.
This orientation is consistent with the preference of a high inclination angle
and a small position angle for the same RIAF model,
constrained by size measurements in 7mm and 3mm \citep{Huang07}.
However, our Model A with the same combination is excluded by
observational data because it predicts too compact an image for the emission region.
This is due to our use of the plasma wave heating mechanism.

There are other efforts of estimations on the black hole parameters,
especially orientation, based on different observations and
theoretical models. E.g., \citet{MBF07} derived a high inclination
angle and a favorite value of $\sim 105^\circ$ for position angle,
by comparing the visibilities observed in 7mm and predicted by a
jet-dominated model; \citet{Meye07} derived a high inclination angle
and a range of $\sim 60^\circ - 108^\circ$ for position angle, by
assuming that an orbiting blob contributes to near-infrared flares
and a jet component contributes to mm-to-submm emission. These
models describe morphologies at 1.3mm wavelength totally different
from an accretion flow, which can only be distinguished by the
future direct imaging.

\subsection{Evaluations On The $(u,v)$ Coverage}

In order to obtain a good $(u,v)$ coverage, we need simultanesous
observation on baselines with various lengths and directions.
According to theoretical predictions, baseline lengths of $\lesssim
3{\rm G}\lambda$ are very important because the valley point
structure of the visibility profile may be detected. Fortunately,
there are five potential stations, Hawaii (including JCMT), CARMA,
SMT, LMT, and the Chilean (including ALMA), can provide such
suitable baseline lengths. Although the coverage is still poor, they
can perform simultaneous observations at similar lengths but at
almost perpendicular directions. The basic geometry of the emission
region of Sgr A* may be estimated by them. Furthermore, ALMA
provides excellent sensitivity on baselines as long as $\sim 7{\rm
G}\lambda$ (on baseline AH). It is expected to make a critical
contribution to future sub-mm VLBI observations of Sgr A*.

The proposed station in New Zealand, MT-COOK, also makes great
contribution. It provides long baselines at a large range of
direction. With its participation, these eight stations cover almost
all the directions at lengths of $\gtrsim 5{\rm G}\lambda$. The
longest baseline length of $\sim 8{\rm G}\lambda$, which can provide
the highest resolution, is also achieved by MT-COOK (on baselines CM
and MS).

With the Earth's rotation, these eight stations produce sufficient
$(u,v)$ coverage to produce images of Sgr A*. However, it takes
almost 11 hours to complete. In this paper, we treat observations
within 2 hours as the simultaneous and those within 4 hours as the
quasi-simultaneous. The estimate of the basic geometry of the
emission region is made from the hypothetic observation of four
hours duration, e.g., with \textit{sub-coverage i} mentioned in
Sec.\ref{cov} adopted. Thus, we have to assume that Sgr A* is in its
stable state during the entire 4-hour period. If we want to make
more detailed estimates using coverage in a larger range of
directions, we must assume that this stable state exists even
longer. However, a number of flares have been observed in Sgr A* at
1.3mm, on timescales as short as 30 minutes. Therefore, reliable
estimates can only be made when either Sgr A* is quiescent, the
flaring region is subdominant at 1.3mm, or the observation lasts for
$\lesssim 1$ hour (resulting in poorer coverage and less opportunity
to detect the visibility valley associated with the black hole
shadow).

There is one more short-coming in the $(u,v)$ coverage.
Region with baseline lengths of $\gtrsim 5{\rm G}\lambda$ is covered in
almost all the directions. However, the most important region with
baseline lengths of $\gtrsim 3{\rm G}\lambda$ is lack of coverage
in northeast-southwest direction. During any simultaneous observation
within 2 hours, coverage in two perpendicular directions can be
obtained. However, at least one more direction between those two is
necessary for a good estimation of the emission region geometry, also
for a higher possibility of detection on the valley point structure.
This can only be improved with an additional new station included, e.g., one in New Zealand.

\acknowledgments

This work was supported in part by the National Natural
Science Foundation of China (grants 10573029, 10625314, 10633010,
and 10821302) and the Knowledge Innovation Program of the
Chinese Academy of Sciences (Grant No. KJCX2-YW-T03), and sponsored
by the Program of Shanghai Subject Chief Scientist (06XD14024) and
the National Key Basic Research Development Program of China
(No.2007CB815405). 
L.H. is supported by China Postdoctoral Science Foundation 
(grant 20090450822). 
We are grateful to Dr. S. Liu, Dr. M. J. Cai, and Prof. R. E.
Taam for comments on the modelling, to Prof. P. T. P. Ho for suggestions in visibility
analysis, and to Dr. K. Asada for providing the uv-coverage.


\clearpage

\end{document}